\documentclass[
	envcountsect, 
	twocolumn
	]{svjour3}

\usepackage[utf8]{inputenc}

\usepackage{newtxtext}

\usepackage{amsmath}

\usepackage{microtype}

\usepackage{graphicx}

\usepackage{xspace}

\usepackage{units}

\usepackage{todonotes}
\usepackage{color}

\newenvironment{magentaEnv}{\color{magenta}}{}
\newcommand{\todoDiss}[1]{}

\usepackage{booktabs}

\usepackage{subfigure}

\usepackage{hyperref}

\usepackage{cleveref}
\crefrangeformat{equation}{(#3#1#4) to~(#5#2#6)}
\crefmultiformat{equation}{(#2#1#3)}{ and~(#2#1#3)}{, (#2#1#3)}{ and~(#2#1#3)}

\usepackage[numbers,sort&compress]{natbib}

\usepackage{rotating}

\usepackage{xfp}

\usepackage{tikz}
\usepackage{pgfplots}
\pgfplotsset{compat=1.16}

\newcommand{\includegraphicsdpi}[2]{\includegraphics[scale=\fpeval{72/#1}]{#2}}%

\usepackage{soul}

\usepackage{etoolbox}
\usepackage{xifthen}
\usepackage{bm}
\usepackage{amssymb}
\usepackage{mathtools}

\newcommand{\ie}{i.e.,\xspace}
\newcommand{\eg}{e.g.,\xspace}
\newcommand{\cf}{cf.~}

\newcommand{\sutsu}{surface-to-surface\xspace}

\newcommand{\btb}{beam-to-beam\xspace}

\newcommand{\btv}{beam-to-volume\xspace}

\newcommand{\bts}{beam-to-solid\xspace}

\newcommand{\btsvc}{BTV-POS\xspace}
\newcommand{\btsvFull}{beam-to-volume\xspace}

\newcommand{\btsvcX}{BTV-POS-X\xspace}
\newcommand{\BtsvcXFull}{Extended beam-to-volume coupling\xspace}
\newcommand{\btsvr}{BTV-ROT\xspace}
\newcommand{\btsvrc}{BTV-FULL\xspace}

\newcommand{\btss}{beam-to-surface\xspace}
\newcommand{\btsso}{beam-to-solid-surface\xspace}
\newcommand{\btssh}{beam-to-shell-surface\xspace}
\newcommand{\btssLong}{beam-to-surface\xspace}
\newcommand{\btssc}{beam-to-surface coupling\xspace}
\newcommand{\BtsscFull}{Beam-to-surface coupling\xspace}
\newcommand{\btssfull}{BTS-FULL\xspace}
\newcommand{\btssfullcons}{BTS-FULL-CONS\xspace}
\newcommand{\btssfullref}{BTS-FULL-REF\xspace}
\newcommand{\btssfulldisp}{BTS-FULL-DISP\xspace}
\newcommand{\btsstrans}{BTS-POS\xspace}
\newcommand{\btssrot}{BTS-ROT\xspace}

\newcommand{\sr}{Simo--Reissner\xspace}
\newcommand{\kl}{Kirchhoff--Love\xspace}

\newcommand{\cs}[1][]{cross-section#1\xspace}

\newcommand{\gale}{Gauss--Legendre\xspace}

\newcommand{\boltz}{Boltzmann\xspace}
\newcommand{\nr}{Newton--Raphson\xspace}
\newcommand{\cosserat}{Cosserat\xspace}
\newcommand{\nurbs}{NURBS\xspace}

\newcommand{\gexact}{geometrically exact\xspace}
\newcommand{\Gexact}{Geometrically exact\xspace}

\renewcommand{\hex}[1]{\emph{hex#1}\xspace}
\newcommand{\tet}[1]{\emph{tet#1}\xspace}

\newcommand{\fad}{FAD\xspace}

\newcommand{\R}[1]{\mathbb{R}^{#1}}
\newcommand{\C}[1]{\ensuremath{C^{#1}}}
\newcommand{\SO}{SO^3}

\newcommand{\infsup}{\mathrm{inf}\text{-}\mathrm{sup}}

\newcommand{\e}[1]{\tns{e}_{#1}}
\newcommand{\ex}{\e{1}}
\newcommand{\ey}{\e{2}}
\newcommand{\ez}{\e{3}}

\newcommand{\pfrac}[2]{\frac{\partial #1}{\partial #2}}
\newcommand{\pfracinline}[2]{\partial #1 / \partial #2}

\DeclareMathOperator{\lin}{Lin}

\newcommand{\tn}[2]{%
\ifnumcomp{#1}{=}{1}{\underline{\boldsymbol{#2}}}
{\ifnumcomp{#1}{=}{2}{\underline{\boldsymbol{#2}}}
{\ifnumcomp{#1}{>}{2}{Higher order tensor not yet implemented!}
{Wrong tensor order given}%
}}}
\newcommand{\tns}[1]{\tn{1}{#1}}
\newcommand{\tnss}[1]{\tn{2}{#1}}
\newcommand{\tnsO}{\tns{0}} 
\newcommand{\tnssI}{\tnss{I}} 

\newcommand{\vv}[1]{\boldsymbol{\mathsf{#1}}}
\newcommand{\mat}[1]{\boldsymbol{\mathsf{#1}}}

\newcommand{\vectO}{\vv{0}}
\newcommand{\matO}{\mat{0}}
\newcommand{\matI}{\mat{I}}
\newcommand{\norm}[1]{\left\|#1\right\|}
\newcommand{\normalize}[1]{\frac{#1}{\norm{#1}}}
\newcommand{\relError}[2]{\frac{\norm{#1 - #2}}{\norm{#2}}}

\newcommand{\tr}{^{\mathrm T}}

\newcommand{\inv}{^{-1}}

\renewcommand{\vector}[1]{\begin{bmatrix}#1\end{bmatrix}}
\renewcommand{\matrix}[1]{\begin{bmatrix}#1\end{bmatrix}}

\newcommand{\brackets}[4][]{%
\ifthenelse{\isempty{#1}}{%
\left#2#4\right#3%
}{%
\ifnumcomp{#1}{=}{0}{#2#4#3}
{\ifnumcomp{#1}{=}{1}{\bigl#2#4\bigr#3}
{\ifnumcomp{#1}{=}{2}{\Bigl#2#4\Bigr#3}
{\ifnumcomp{#1}{=}{3}{\biggl#2#4\biggr#3}
{\ifnumcomp{#1}{=}{4}{\Biggl#2#4\Biggr#3}
{size not supported}
}}}}}}

\newcommand{\br}[2][]{\brackets[#1]{(}{)}{#2}}

\newcommand{\at}[2][]{\brackets[#1]{.}{|}{#2}}

\newcommand{\encapsulate}[1]{{#1}}

\newcommand{\placeholder}{(\cdot)}

\newcommand{\letterbeam}{B}
\newcommand{\lettersolid}{S}
\newcommand{\lettervolume}{V}
\newcommand{\lettersurface}{S}

\newcommand{\letterbeamcenterline}{r}
\newcommand{\letterbeampos}{\letterbeamcenterline}
\newcommand{\letterbeamrot}{\theta}

\newcommand{\Esolid}{E_\lettersolid}
\newcommand{\Ebeam}{E_\letterbeam}
\newcommand{\nusolid}{\nu_\lettersolid}
\newcommand{\nubeam}{\nu_\letterbeam}
\newcommand{\radius}{R}

\newcommand{\esolidName}{e}
\newcommand{\esolid}{{(\esolidName)}}

\newcommand{\sbeam}{s}

\newcommand{\domainSolidref}{\Omega_{\lettervolume,0}}
\newcommand{\domainSolid}{\Omega_\lettervolume}

\newcommand{\domainSolidSurfaceref}{\partial\domainSolidref}
\newcommand{\domainSolidSurface}{\partial\domainSolid}
\newcommand{\domainSolidNeumann}{\Gamma_\sigma}
\newcommand{\domainSurfaceref}{\Omega_{\lettersurface,0}}
\newcommand{\domainSurface}{\Omega_\lettersurface}

\newcommand{\domainBeamref}[1][]{\Omega_{\letterbeam#1,0}}
\newcommand{\domainBeam}[1][]{\Omega_{\letterbeam#1}}

\newcommand{\domainCoupling}{\Gamma_{c}}
\newcommand{\domainCouplingh}{\Gamma_{c,h}}

\newcommand{\intSolid}[1]{\int_{\domainSolidref}{ #1 \,\mathrm d V_0}\,}
\newcommand{\intSolidShellSurface}[1]{\int_{\domainSurfaceref}{ #1 \,\mathrm d A_0}\,}
\newcommand{\intSolidNeumann}[1]{\int_{\domainSolidNeumann}{ #1 \,\mathrm d A_0}\,}
\newcommand{\intBeamCenterline}[1]{\int_{\domainBeamref}{ #1 \,\mathrm d \sbeam}\,}
\newcommand{\intCoupling}[1]{\int_{\domainCoupling}{ #1 \,\mathrm d \sbeam}\,}
\newcommand{\intCouplingh}[1]{\int_{\domainCouplingh}{ #1 \,\mathrm d \sbeam}\,}

\newcommand{\nsurface}{n_{\lettersurface}}
\newcommand{\nbeam}{n_B}
\newcommand{\nlagrange}{n_\lambda}
\newcommand{\indexsolid}{k}
\newcommand{\indexbeam}{l}
\newcommand{\indexlagrange}{j}

\newcommand{\sumsurface}[1]{\sum_{\indexsolid=1}^{\nsurface}{#1}}
\newcommand{\sumbeam}[1]{\sum_{\indexbeam=1}^{\nbeam}{#1}}
\newcommand{\sumlagrange}[1]{\sum_{\indexlagrange=1}^{\nlagrange}{#1}}

\newcommand{\nameinternal}{\text{int}}
\newcommand{\nameexternal}{\text{ext}}

\newcommand{\rotMat}{\tnss{R}}

\DeclareMathOperator{\rv}{rv}

\newcommand{\triad}{\tnss{\Lambda}}
\newcommand{\triadbeam}{\triad_{\letterbeam}}
\newcommand{\triadbeamO}{\triad_{\letterbeam,0}}

\newcommand{\gtriad}[1]{\tns{g}_{#1}}

\newcommand{\gtriadbeam}[1]{\tns{g}_{\letterbeam#1}}
\newcommand{\gtriadbeamO}[1]{\tns{g}_{\letterbeam#1,0}}

\newcommand{\gtriadtilde}[1]{\tilde{\tns{g}}_{#1}}

\newcommand{\rotvec}{\tns{\psi}}
\newcommand{\rotvecaxis}{\tns{e}_{\psi}}
\newcommand{\rotvecnorm}{\psi}

\newcommand{\rotvecbeam}{\rotvec_{\letterbeam}}

\newcommand{\rotvecbeamsolid}{\rotvec_{\lettersolid\letterbeam}}

\newcommand{\Sskew}[1][]{\tnss{S}\ifthenelse{\isempty{#1}}{}{\br{#1}}}

\newcommand{\penRot}{\epsilon_{\letterbeamrot}}

\newcommand{\dcouplingPotentialMortar}{\delta\Pi_\lambda}

\newcommand{\posMortarName}{\lambda_\letterbeamcenterline}
\newcommand{\couplingPotentialPosMortar}{\Pi_{\posMortarName}}
\newcommand{\dcouplingPotentialPosMortar}{\delta\couplingPotentialPosMortar}
\newcommand{\dcouplingPotentialPosMortarh}{\delta\Pi_{{\posMortarName},h}}

\newcommand{\rotMortarName}{\lambda_\letterbeamrot}
\newcommand{\couplingPotentialRotMortar}{\Pi_{\rotMortarName}}

\newcommand{\dcouplingPotentialRotMortar}{\delta\couplingPotentialRotMortar}

\newcommand{\dWposLambda}{\delta W_{\posMortarName}}
\newcommand{\dWposLambdaSurfRef}{\delta W_{\posMortarName}^\text{\nameReferenceShort}}
\newcommand{\dWposLambdaSurfDisp}{\delta W_{\posMortarName}^\text{\nameDisplacementShort}}
\newcommand{\dWposLambdah}{\delta W_{\posMortarName,h}}

\newcommand{\dWposC}{\delta W_{C_\letterbeamcenterline}}
\newcommand{\dWposCSurfRef}{\delta W_{C_\letterbeamcenterline}^\text{\nameReferenceShort}}
\newcommand{\dWposCSurfDisp}{\delta W_{C_\letterbeamcenterline}^\text{\nameDisplacementShort}}
\newcommand{\dWposCh}{\delta W_{C_\letterbeamcenterline,h}}

\newcommand{\DQbeamrot}{\Delta\vv{d}^\letterbeam_{\letterbeamrot}}
\newcommand{\DQbeamPlace}{\Delta\vv{d}^\letterbeam_{\placeholder}}
\newcommand{\RcsolidPos}{\vv{r}_{c,\posMortarName}^\lettersolid}

\newcommand{\RcbeamPos}{\vv{r}_{c,\posMortarName}^\letterbeam}

\newcommand{\RcPos}{\vv{r}_{c,\posMortarName}}

\newcommand{\RcsolidRot}{\vv{r}_{c,\rotMortarName}^\lettersolid}

\newcommand{\RcbeamRot}{\vv{r}_{c,\rotMortarName}^\letterbeam}

\newcommand{\RcRot}{\vv{r}_{c,\rotMortarName}}

\newcommand{\QcPlace}{\mat{Q}_{\placeholder\placeholder}^{\text{ROT}}}

\newcommand{\Qcss}{\mat{Q}_{ss}^{\text{ROT}}}
\newcommand{\Qcsb}{\mat{Q}_{s\letterbeamrot}^{\text{ROT}}}
\newcommand{\Qcsl}{\mat{Q}_{s\rotMortarName}^{\text{ROT}}}

\newcommand{\Qcbs}{\mat{Q}_{\letterbeamrot s}^{\text{ROT}}}
\newcommand{\Qcbb}{\mat{Q}_{\letterbeamrot\letterbeamrot}^{\text{ROT}}}
\newcommand{\Qcbl}{\mat{Q}_{\letterbeamrot\rotMortarName}^{\text{ROT}}}

\newcommand{\Qcls}{\mat{Q}_{\rotMortarName s}^{\text{ROT}}}
\newcommand{\Qclb}{\mat{Q}_{\rotMortarName\letterbeamrot}^{\text{ROT}}}

\newcommand{\ScalingMatrixRot}{\mat{V}_{\rotMortarName}}

\newcommand{\lagrangePos}{\tns{\lambda}_{\letterbeamcenterline}}
\newcommand{\lagrangePosh}{\tns{\lambda}_{\letterbeamcenterline,h}}

\newcommand{\dlagrangePos}{\delta\lagrangePos}

\newcommand{\QlagrangePos}{\vv{\lambda}_{\letterbeamcenterline}}
\newcommand{\dQlagrangePos}{\delta\QlagrangePos}

\newcommand{\qlagrangePosn}{\tns{\lambda}_{\letterbeamcenterline,\indexlagrange}}

\newcommand{\dqlagrangePosn}{\delta\qlagrangePosn}

\newcommand{\NlagrangePosni}{\Phi_{\letterbeamcenterline,j}}

\newcommand{\QlagrangeRot}{\vv{\lambda}_{\letterbeamrot}}

\newcommand{\dWsolid}{\delta W^\lettersolid}

\newcommand{\Xsolid}{\tns{X}_\lettersolid}
\newcommand{\xsolid}{\tns{x}_\lettersolid}
\newcommand{\dxsolid}{\delta\tns{x}_\lettersolid}
\newcommand{\usolid}{\tns{u}_\lettersolid}

\newcommand{\dusolid}{\delta\usolid}
\newcommand{\F}{\tnss{F}}
\newcommand{\Spk}{\tnss{S}}
\newcommand{\E}{\tnss{E}}
\newcommand{\dE}{\delta \E}
\newcommand{\loadSolidBody}{\hat{\tns{b}}}
\newcommand{\loadSolidSurface}{\hat{\tns{t}}}
\newcommand{\loadShellBody}{\hat{\tns{f}}}
\newcommand{\shellStressTensor}{\tnss{n}}
\newcommand{\dshellStrain}{\delta\tnss{\varepsilon}}
\newcommand{\shellBendingMomentTensor}{\tnss{m}}
\newcommand{\dshellCurvature}{\delta\tnss{\kappa}}
\newcommand{\Fsurface}{\tnss{F}_\lettersurface}

\newcommand{\Xsolidh}{\tns{X}^\lettersolid_h}
\newcommand{\usolidh}{\tns{u}^\lettersolid_h}

\newcommand{\dusolidh}{\delta\tns{u}^{\lettersolid}_h}

\newcommand{\Qsolid}{\encapsulate{{\vv{d}^\lettersolid}}}
\newcommand{\dQsolid}{\delta\Qsolid}
\newcommand{\DQsolid}{\Delta\Qsolid}
\newcommand{\qsolid}{\encapsulate{{\vv{d}^\lettersolid}}}

\newcommand{\qsolidn}{\encapsulate{\tns{d}^{\lettersolid}_\indexsolid}}
\newcommand{\qxsolid}{\encapsulate{\vv{x}^{\lettersolid}}}

\newcommand{\qxsolidn}{\encapsulate{\tns{x}^{\lettersolid}_\indexsolid}}
\newcommand{\Qxsolid}{\encapsulate{\vv{x}^{\lettersolid}}}

\newcommand{\qXsolidn}{\encapsulate{\tns{X}^{\lettersolid}_\indexsolid}}
\newcommand{\QXsolid}{\encapsulate{\vv{X}^{\lettersolid}}}

\newcommand{\dqsolidn}{\encapsulate{\delta\qsolidn}}

\newcommand{\Nsolidn}{N_\indexsolid}

\newcommand{\Kss}{\mat{K}^\lettersolid_{ss}}
\newcommand{\Rsolid}{\vv{r}^\lettersolid}

\newcommand{\xisurface}{\xi^\lettersolid}
\newcommand{\etasurface}{\eta^\lettersolid}

\newcommand{\xisurfaceCpp}{\xisurface_c}
\newcommand{\etasurfaceCpp}{\etasurface_c}

\newcommand{\Pbeam}{\Pi_{\nameinternal,\letterbeam}}
\newcommand{\dPbeam}{\delta \Pbeam}
\newcommand{\dWbeam}{\delta W^\letterbeam}

\newcommand{\dWbeamext}{\dWbeam_{\nameexternal}}

\newcommand{\rbeamO}{\tns{r}_0}

\newcommand{\rbeam}{\tns{r}}

\newcommand{\drbeam}{\delta\rbeam}
\newcommand{\drbeamh}{\delta\rbeam_h}
\newcommand{\ubeam}{\tns{u}_\letterbeam}
\newcommand{\dubeam}{\delta\ubeam}

\newcommand{\rbeamh}{\tns{r}_{h}}

\newcommand{\Qbeam}{\vv{d}^\letterbeam}
\newcommand{\dQbeam}{\encapsulate{\delta\Qbeam_r}}
\newcommand{\DQbeam}{\encapsulate{\Delta\Qbeam_r}}

\newcommand{\qbeamn}{\vv{d}^{\letterbeam}_\indexbeam}
\newcommand{\qbeamer}{\tns{d}^{B,r}_{\indexbeam}}
\newcommand{\qbeamet}{\tns{d}^{B,t}_{\indexbeam}}
\newcommand{\dqbeamer}{\delta\qbeamer}
\newcommand{\dqbeamet}{\delta\qbeamet}
\newcommand{\qxbeamer}{\encapsulate{\tns{r}^{B}_{\indexbeam}}}
\newcommand{\qxbeamet}{\encapsulate{\tns{t}^{B}_{\indexbeam}}}

\newcommand{\dqbeamn}{\encapsulate{\delta\qbeamn}}
\newcommand{\qxbeam}{\vv{x}^{\letterbeam}}

\newcommand{\qxbeamn}{\encapsulate{\vv{x}^{\letterbeam}_\indexbeam}}
\newcommand{\Qxbeam}{\vv{x}^{\letterbeam}}

\newcommand{\QXbeam}{\vv{X}^{\letterbeam}}

\newcommand{\qxbeamrot}{\encapsulate{\tns{\psi}^{\letterbeam}_{\indexbeam}}}
\newcommand{\qxbeamrotn}{\encapsulate{\tns{\psi}^{\letterbeam}_{\indexbeam,i}}}
\newcommand{\qxbeamrotnp}{\encapsulate{\tns{\psi}^{\letterbeam}_{\indexbeam,i+1}}}
\newcommand{\qxbeamrotincrement}{\encapsulate{\Delta\tns{\theta}^{\letterbeam}_{\indexbeam,i}}}

\newcommand{\Nbeamn}{\mat{H}_\indexbeam}
\newcommand{\Nbeamr}{H_\indexbeam^r}
\newcommand{\Nbeamt}{H_\indexbeam^t}

\newcommand{\xibeam}{\xi^\letterbeam}
\newcommand{\xibeampaper}{\sbeam}

\newcommand{\Krr}{\mat{K}^\letterbeam_{rr}}
\newcommand{\Krt}{\mat{K}^\letterbeam_{r\letterbeamrot}}
\newcommand{\Ktr}{\mat{K}^\letterbeam_{\letterbeamrot r}}
\newcommand{\Ktt}{\mat{K}^\letterbeam_{\letterbeamrot\letterbeamrot}}
\newcommand{\KbeamPlace}{\mat{K}^\letterbeam_{\placeholder\placeholder}}
\newcommand{\Rbeam}{\vv{r}^\letterbeam_{r}}
\newcommand{\RbeamPos}{\vv{r}^\letterbeam_{\letterbeampos}}
\newcommand{\RbeamRot}{\vv{r}^\letterbeam_{\letterbeamrot}}
\newcommand{\RbeamPlace}{\vv{r}^\letterbeam_{\placeholder}}

\newcommand{\srTension}{\tns{\Gamma}}
\newcommand{\srTensionC}{\tnss{C}_F}
\newcommand{\srBending}{\tns{\Omega}}
\newcommand{\srBendingC}{\tnss{C}_M}

\newcommand{\penPos}{\epsilon_\letterbeamcenterline}

\newcommand{\M}{\mat{M}}
\newcommand{\D}{\mat{D}}

\newcommand{\Mn}{\encapsulate{\mat{M}^{(\indexlagrange,\indexsolid)}}}
\newcommand{\Dn}{\encapsulate{\mat{D}^{(\indexlagrange,\indexbeam)}}}
\newcommand{\scalingMatrix}{\mat{\kappa}_{r}}
\newcommand{\ScalingMatrix}{\mat{V}_{r}}

\newcommand{\nameSurf}{\text{surf}}
\newcommand{\nameVolume}{\text{vol}}

\newcommand{\surfaceNormalO}{\tns{N}}
\newcommand{\surfaceNormal}{\tns{n}}
\newcommand{\dsurfaceNormal}{\delta\tns{n}}
\newcommand{\surfaceNormalDistance}{d}
\newcommand{\surfaceNormalDistanceUL}{\surfaceNormalDistance_{c}}
\newcommand{\surfaceNormalDistanceULO}{\surfaceNormalDistance_{c,0}}

\newcommand{\surfaceNormalh}{\tns{n}_{h}}

\newcommand{\dsurfaceNormalh}{\delta\surfaceNormalh}

\newcommand{\dxsoliddxi}{\tns{x}_{S,\xisurface}}
\newcommand{\dxsoliddeta}{\tns{x}_{S,\etasurface}}
\newcommand{\dXsoliddxi}{\tns{X}_{S,\xisurface}}
\newcommand{\dXsoliddeta}{\tns{X}_{S,\etasurface}}

\newcommand{\triadVolume}{\triad_{\nameVolume}}
\newcommand{\triadsurface}{\triad_{\nameSurf}}
\newcommand{\triadsurfaceO}{\triad_{\nameSurf,0}}
\newcommand{\triadsurfaceRef}{\tilde{\triad}_{\nameSurf}}
\newcommand{\triadsurfaceRel}{\tilde{\triad}_{\nameSurf,0}}

\newcommand{\qsurfg}{\vv{q}}

\newcommand{\qsurfn}{\encapsulate{\qsurfg^{(\indexlagrange)}}}
\newcommand{\Qsurfg}{\mat{Q}}

\newcommand{\Qsurfn}{\encapsulate{\Qsurfg^{(\indexlagrange,\indexsolid)}}}
\newcommand{\Qsurfsurfg}{\mat{Q}_{ss}}

\newcommand{\nameConsistentShort}{CONS}
\newcommand{\nameConsistent}{\btsstrans-\nameConsistentShort\xspace}
\newcommand{\nameReferenceShort}{REF}
\newcommand{\nameReference}{\btsstrans-\nameReferenceShort\xspace}
\newcommand{\nameDisplacementShort}{DISP}
\newcommand{\nameDisplacement}{\btsstrans-\nameDisplacementShort\xspace}

\newcommand{\nodesurface}{k}
\newcommand{\nodeNormalAverage}{\tns{n}_{\text{AVG},\nodesurface}}
\newcommand{\nodeNormal}{\tns{n}^{\esolid}_{\nodesurface}}
\newcommand{\nodeNormalNumber}{n_{\text{adj},\nodesurface}}

\newcommand{\virtDisp}{\delta\tns{u}}
\newcommand{\virtRot}{\delta\tns{\phi}}

\newcommand{\thickness}{t}

\journalname{}

\begin{document}

\title{%
A consistent mixed-dimensional coupling approach for 1D \cosserat beams and 2D surfaces in 3D space%
}

\author{
    Ivo Steinbrecher\and
    Nora Hagmeyer\and
    Christoph Meier\and
    Alexander Popp
}

\institute{
    I. Steinbrecher, N. Hagmeyer, A. Popp\at
    Institute for Mathematics and Computer-Based Simulation (IMCS),\\
    University of the Bundeswehr Munich,\\
    Werner-Heisenberg-Weg 39, D-85577 Neubiberg, Germany\\
    \email{ivo.steinbrecher@unibw.de}
    \and
    C. Meier\at
    Simulation of Additive Manufacturing Processes (SAM),\\
    Technical University of Munich,\\
    Freisinger Landstraße 52, D-85748 Garching b. München, Germany
}

\date{Received: date / Accepted: date}

\maketitle

\begin{abstract}
The present article proposes a novel computational method for coupling arbitrarily curved 1D fibers with a 2D surface as defined, \eg by the 2D surfaces of a 3D solid body or by 2D shell formulations.
The fibers are modeled as 1D \cosserat continua (beams) with six local degrees of freedom, three positional and three rotational ones.
A kinematically consistent 1D-2D coupling scheme for this problem type is proposed considering the positional and rotational degrees of freedom along the beams.
The positional degrees of freedom are coupled by enforcing a constant normal distance between a point on the beam centerline and a corresponding point on the surface.
This strategy requires a consistent description of the surface normal vector field to guarantee fundamental mechanical properties such as conservation of angular momentum.
Coupling of the rotational degrees of freedom of the beams and a suitable rotation tensor representing the local orientation within a solid volume has been considered in a previous contribution.
In the present work, this coupling approach will be extended by constructing rotation tensors that are representative of local surface orientations.
Several numerical examples demonstrate the consistency, robustness and accuracy of the proposed method.
To showcase its applicability to multi-physics systems of practical relevance, the fluid-structure interaction example of a vascular stent is presented.
\end{abstract}

\section{Introduction}
\label{sec:introduction}

Compound systems composed of slender one-dimensional (1D) components, \ie where one spatial dimension is significantly larger than the other two, interconnected with higher-dimensional continua can be found in a variety of different fields.
Applications for this class of problems include civil engineering, where steel girders are used to support concrete slabs, or in mechanical engineering, where lightweight structures are realized by stabilizing thin shells with struts.
Leaving the realm of classical engineering applications, similar principles are also employed in biomedical systems, such as the interaction between a stent and its encasing fabric (graft) in endovascular aneurysm repair.
Numerical simulation of such applications is of high importance during the development and design phase to accurately predict and control the desired system behavior.

\begin{figure}
\centering
\subfigure[]{\label{fig:introduction:problem_types:solid}\includegraphics[scale=1]{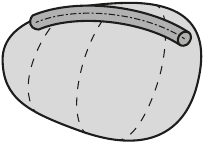}}
\hfill
\subfigure[]{\label{fig:introduction:problem_types:shell}\includegraphics[scale=1]{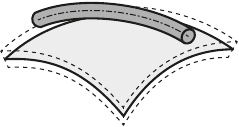}}
\caption{Types of considered \btss scenarios in this work, \subref{fig:introduction:problem_types:solid} the beam is coupled to the 2D surface of a classical 3D \boltz continua (\btsso scenario) and \subref{fig:introduction:problem_types:shell} the beam is coupled to a reduced dimensional 2D shell formulation (\btssh scenario).}
\label{fig:introduction:problem_types}
\end{figure}

From a geometrical point of view, the problem considered in the present work consists of 1D beams coupled to two-dimensional (2D) surfaces.
Thus, the resulting problem can be classified as a mixed-dimensional 1D-2D coupling problem.
This will be denoted as a \emph{\btssLong} coupling problem throughout this contribution.
Two different surface types are considered in the present contribution: 2D surfaces of classical 3D \boltz continua and reduced dimensional 2D shell formulations, \cf \Cref{fig:introduction:problem_types}.
The proposed coupling formulations are exactly the same for both types of surfaces.
However, some theoretical and numerical aspects specifically apply to only one of the two variants, in which case we will use the terminology \emph{\btsso} and \emph{\btssh}.
The term \emph{structure} will be used to denote the actual structure to which the coupling surface belongs, independent of its origins in either a 3D solid continuum or a 2D shell formulation.

The present modeling approach for the \btss coupling problem employs accurate and efficient 1D models for the beam-like structures based on \gexact beam theory~\cite{Reissner1972, Simo1986a, Simo1986b, Cardona1988, Ibrahimbegovic1995, Crisfield1999, Romero2004, Meier2019, Betsch2002, Sonneville2014}.
The beams are represented by 1D curves in 3D space, \ie the beam centerline that connects the centroids of the beam \cs[s].
Each point along the beam centerline has six degrees of freedom (three positional and three rotational ones), \ie the beam model can be identified as a 1D \cosserat continuum.
The resulting \btss coupling problem has two desirable features, \cf \cite{Steinbrecher2020, Steinbrecher2022}: (i) Both the structure (\ie 3D solid or 2D shell) and the beam can be modeled and discretized individually.
Therefore, well-established discretization schemes for the structure and the beam can be used without modifications.
(ii) Employing 1D beam models results in computationally efficient finite element discretizations, which reduces the number of unknowns required to represent the beam-like components by several orders of magnitude as compared to a modeling approach based on 3D continuum theory.
In the literature, mixed-dimensional coupling between structural beam theories and solid continua is often addressed to model fiber-reinforced materials, \eg~\cite{Phillips1976, Chang1987, Elwi1989, Ranjbaran1996, Gomes2001, Kerfriden2020}.
However, in all of the aforementioned works, the coupling is an embedded 1D-3D coupling since the 1D fiber reinforcements are placed inside the solid domain and are directly coupled to the 3D solid volume.
Furthermore, 1D string-like models with a limited representation of the relevant modes of deformation, \ie only axial deformation, were used in these contributions to represent the 1D curves.
Coupling approaches for full beam theories in 1D-3D beam-to-volume coupling problems have been investigated more recently, \eg in \cite{Durville2007, Steinbrecher2020, Steinbrecher2022, Khristenko2021, Firmbach2024}.
Compared to the previously mentioned string models, beam theories contain additional deformations modes, \ie bending, torsion and shear, which allows for a more realistic representation of the nonlinear force-displacement relations caused by the reinforcements.
In \cite{Durville2007} a collocation method is used to couple the beams to the solid.
A mortar-type approach to couple the positional degrees of freedom of the beam centerline to the solid is presented in~\cite{Steinbrecher2020} and a mortar-type approach for \emph{full} coupling, \ie positional \emph{and} rotational coupling, is presented in \cite{Steinbrecher2022}.
In \cite{Khristenko2021} the coupling constraints of a \btsvFull problem are formulated on the beam's surface of the beam and are subsequently projected onto the beam centerline considering a Taylor series expansion of the solid displacement field.
Apart from the 1D-3D coupling problems discussed so far, a truly 1D-2D coupling is presented in \cite{Konyukhov2015}.
However, the surfaces are assumed to be rigid in that contribution, which heavily limits the applicability to real life engineering problems.

In the present work, we propose the first truly mixed-dimensional 1D-2D mortar-type approach for \btss coupling problems.
This is an extension of the authors' previous contributions on positional \btsvFull (\btsvc) and rotational \btsvFull (\btsvr) coupling problems, \cf \cite{Steinbrecher2020,Steinbrecher2022}, to \btss coupling problems.
The transition from a 1D-3D to a 1D-2D mixed-dimensional coupling poses two significant challenges:
(i)~The primary challenge in modeling the 1D-2D positional \btss (\btsstrans) coupling lies in the dependency of the coupling constraints on the surface normal distance, \ie the offset between the beam centerline and the surface.
In the considered \btss application cases this offset is unavoidable and it can even be motivated physically: in many real-life applications, the beams are tied (\eg welded or glued) to the surface.
In such cases, the offset between the beam centerline and the surface is typically equal to the \cs radius of the beam.
(ii)~For rotational \btss (\btssrot) coupling, a suitable surface orientation field is required on the surface.
A detailed discussion on rotation tensors that are representative of the local orientation of a 3D solid continuum is given in \cite{Steinbrecher2022}.
However, directly applying these approaches to the cases considered in this work is not feasible.
This is because our analysis extends beyond 3D solid continua to include 2D shell formulations, and because even for 3D continua, rotation tensors that are representative of the local orientation \emph{inside} a 3D solid continuum come with significant drawbacks when being evaluated on the surface of the 3D continuum.
Therefore, an additional scientific contribution of this work is the construction of a surface orientation suitable for use in \btssrot.
As pointed out in~\cite{Steinbrecher2022}, it is essential to couple positions \emph{and} rotations to achieve a full coupling of all beam deformation modes to the structure.
This combination of \btsstrans and \btssrot will be referred to as \emph{full} \btss (\btssfull) coupling.

The main novelty of the present work is a thorough analysis of how the surface normal distance dependency in \btsstrans affects the mechanical and numerical properties of the coupled system.
In this work, the term \emph{consistent} implies that no simplifications regarding the surface normal distance are introduced in the further derivation of the coupling constraints.
A consistent treatment of the surface normal distance, especially in the discretized problem setting, can become cumbersome, as it requires the evaluation of the surface normal vector and its second derivatives.
Thus, we present two alternatives to the consistent 1D-2D mixed-dimensional coupling constraints, which do not contain the surface normal distance, and compare them with the consistent variant regarding numerical accuracy and fulfillment of fundamental mechanical principles.
As an important scientific contribution of this work, the necessity of a fully consistent treatment of the surface normal distance in the general case of non-matching 1D-2D interfaces is demonstrated.
Only this consistency within the coupling constraints ensures the fulfillment of fundamental mechanical properties and yields physically meaningful results.
In particular, exact conservation of linear and angular momentum is shown for the resulting consistent 1D-2D coupling scheme.
To the authors' best knowledge, this is the first time that exact conservation of angular momentum is shown for a surface coupling scheme with a non-vanishing surface normal distance.
Additionally, the findings presented in this work are not limited to mixed-dimensional coupling problems, but can be directly transferred to \sutsu coupling problems.
In classical \sutsu mesh tying formulations, the continuous surfaces are assumed to be matching, \ie there is no offset in surface normal direction between them.
However, in cases where there is a physical reason for a non-vanishing surface normal distance, \eg in shell-to-shell mesh tying problems or in the case of real-life systems with geometrical manufacturing inaccuracies, existing \sutsu mesh tying approaches would exhibit similar deficiencies to those highlighted in this work concerning the simplified \btss coupling approaches.
Using the proposed consistent variant would allow to resolve these issues also for classical \sutsu mesh tying.
This further underlines the scientific value of the proposed consistent mixed-dimensional \btss coupling approach.

\begin{figure}
\centering
\subfigure[]{\label{fig:introduction:kelvin_flamant:kelvin}\includegraphics[scale=1]{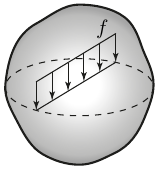}}
\hfil
\subfigure[]{\label{fig:introduction:kelvin_flamant:flamant}\includegraphics[scale=1]{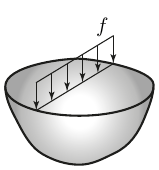}}
\caption{\subref{fig:introduction:kelvin_flamant:kelvin} The Kelvin problem of an embedded line load acting on an infinite solid and \subref{fig:introduction:kelvin_flamant:flamant} the Flamant problem of a line load acting on an infinite solid half space.}
\label{fig:introduction:kelvin_flamant}
\end{figure}

Eventually, it is emphasized that a modeling approach based on mixed-dimensional coupling influences the nature of the underlying mechanical problem.
In the context of embedded 1D-3D coupling this issue has been thoroughly discussed and analyzed for the cases of positional coupling \cite{Steinbrecher2020} and rotational coupling \cite{Steinbrecher2022}.
One of the main consequences is that the analytical solution of the mixed-dimensional coupling problem exhibits a local singularity at the position of the beam centerline.
In the embedded 1D-3D positional coupling case this can be interpreted as a generalization of the well-known Kelvin problem \cite{PodioGuidugli2014,Kelvin1848}, \cf \Cref{fig:introduction:kelvin_flamant:kelvin}, \ie a line load acting on an infinite solid.
The same issue arises in the case of the 1D-2D \btsso scenario, \ie if the considered surface represents the boundary of a 3D volume.
This corresponds to the Flamant problem of a line load acting on an infinite half space~\cite{PodioGuidugli2005}, \cf \Cref{fig:introduction:kelvin_flamant:flamant}.
From a mathematical point of view, the resulting mixed-dimensional coupling formulation is not asymptotically correct as very fine discretizations would approximate the singularity rather than the actual physical problem.
Recently, asymptotically correct mixed-dimensional coupling approaches have been developed, \eg for solid mechanics~\cite{Khristenko2021}, Laplace's equation~\cite{Heltai2023} or fluid-structure-interaction~\cite{Lespagnol2024}.
However, for the 1D-3D \btsvFull case, it has been shown in detail that a non-asymptotically correct coupling formulation does not affect the applicability of the coupling method for the envisioned range of practically relevant discretization resolutions, \ie solid element sizes in the range of the beam \cs diameter or above, \cf\cite{Steinbrecher2020,Steinbrecher2022}.
Similarly, for the proposed 1D-2D \btsso scenario, the lack of asymptotic correctness does not compromise applicability for surface element sizes larger than the beam \cs dimensions.
It should be mentioned that in the \btssh scenario, where the 2D surface corresponds to a \kl-type shell formulation, the aformentioned singularity does not occur.

\todoDiss{Introducing a mixed-dimensional coupling changes the underlying mechanical problem.
In the context of embedded 1D-3D coupling this has been thoroughly discussed and analyzed for positional coupling \cite{Steinbrecher2020} and rotational coupling \cite{Steinbrecher2022}.
One of the main effects is that the analytical solution of the mixed-dimensional coupling problem has a singularity at the beam centerline.
In embedded 1D-3D positional coupling case this can be interpreted as a generalization of the well-known Kelvin problem \cite{PodioGuidugli2014,Kelvin1848}, \cf \Cref{fig:introduction:kelvin_flamant:kelvin}, \ie a line load acting on an infinite solid.
The singularity in the analytical solution prevents the finite element discretization of the mixed-dimensional coupling method to achieve spatial mesh convergence in the fine mesh limit.
However, it has been shown successfully that for mesh sizes in the range of the beam \cs dimensions or larger, which are well within the range of real life engineering applications, the mortar-type 1D-3D positional \emph{and} rotational coupling methods converge towards the exact solution, \ie the one if the beam were modeled with continuum elements.
This is a crucial property for the applicability of mixed-dimensional \bts coupling problems.
The same issue arises in the considered case of 1D-2D \btss coupling, as the analytical solution also has a singularity at the beam centerline, \ie the Flamant problem of a line load acting on an infinite half space \cite{Flamant1892, PodioGuidugli2005}, \cf \Cref{fig:introduction:kelvin_flamant:flamant}.
However, as in the \btsv case this does not impact the applicability of the proposed \btss coupling method, as the singularity of the mixed dimensional coupling problem does not come into effect for the envisioned mesh size relations, \ie solid element sizes that are in the same range as the beam \cs diameter.
For a comprehensive discussion on this topic, the interested reader is referred to \cite{Steinbrecher2020, Steinbrecher2022}.}

The remainder of this work is organized as follows:
In \Cref{sec:problem_formulation}, we state the governing equations for solid and beam formulations as well as for the \btssfull method.
In \Cref{sec:surface_triad}, a suitable procedure for constructing the surface triad is presented to couple the rotations of the surface to the beam \cs orientations.
The finite element discretization of the \btssfull method is presented in \Cref{sec:discretization}.
Furthermore, the construction of a~$\C{0}$-continuous surface normal field for standard Lagrangian finite element interpolations of the surface is elaborated.
Finally, in \Cref{sec:examples}, we present numerical examples for the \btsso as well as the \btssh scenario.
We demonstrate the consistency of the presented \btssfull method, and show comparisons with full 3D continuum approaches, \ie where beam and structure are modeled as 3D continua and discretized by 3D solid finite elements.
The importance of coupling both positions \emph{and} rotations for general \btssc problems is shown.
Furthermore, the applicability of the proposed formulation to real-life engineering and biomedical problems is illustrated.

\section{Problem formulation}
\label{sec:problem_formulation}

\begin{figure*}
\centering
\includegraphics[scale=1]{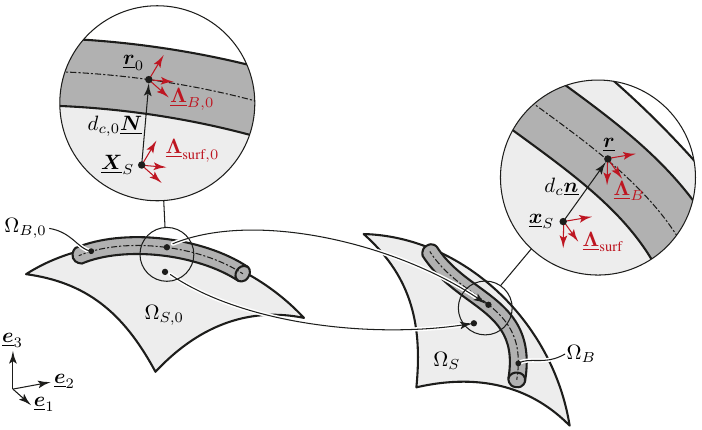}
\caption{Notation for the finite deformation \btssfull coupling problem.}
\label{fig:problem_formulation:btssc_problem}
\end{figure*}

We consider a quasi-static 3D finite deformation \btssfull coupling problem as shown in Figure~\ref{fig:problem_formulation:btssc_problem}.
It is emphasized that the presented \btssfull method is not restricted to quasi-static problems, but can directly be applied to time-dependent problems as well.
A Cartesian frame~$\ex$,~$\ey$ and~$\ez$ serves as fixed frame of reference.
The principle of virtual work (PVW) serves as basis for a subsequent finite element discretization and reads
\begin{equation}
\dWsolid + \dWbeam + \dcouplingPotentialMortar = 0.
\end{equation}
Here, $\dWsolid$ is the total virtual work of the pure structure (\ie 3D solid or 2D shell) problem,~$\dWbeam$ is the total virtual work of the pure beam problem and~$\dcouplingPotentialMortar$ is the virtual work due to coupling forces and moments.
Contributions to the total virtual work of the pure structure and beam problem are independent of the coupling constraints, \ie well-established modeling and discretization techniques can be used for the structure and the beam, \cf \cite{Steinbrecher2020, Steinbrecher2022}.

\subsection{Finite rotations}
\label{sec:finite_rotations}

Before stating the governing equations for the structure, beam and \btssfull coupling problem, a short recap on finite rotations is given here, as a consistent treatment of large rotations is required for the rotational coupling conditions \btssrot.
In \gexact beam theory, the term triad is commonly used to describe the set of three orthonormal vectors defining a beam \cs orientation, \ie
\begin{equation}
\triad = \matrix{\gtriad{1}, \gtriad{2}, \gtriad{3}} \in \SO.
\end{equation}
Here, $\SO$ is the special orthogonal group and $\gtriad{i}$ are the base vectors of the triad.
The triad is equivalent to a rotation tensor, mapping the Cartesian basis vectors $\e{i}$ onto $\gtriad{i}$.
Among others, a triad can be parameterized by the rotation (pseudo-)vector~$\rotvec$, \ie $\triad = \triad(\rotvec)$.
The rotation vector describes a rotation by an angle $\rotvecnorm = \norm{\rotvec}$ around the rotation axis~$\rotvecaxis = \rotvec / \norm{\rotvec}$.
The parametrization can be evaluated with the well-known Rodrigues formula \cite{Argyris1982}
\begin{equation}
\begin{split}
\triad(\rotvec) &= \exp\br{\Sskew[\rotvec]} \\
&= \tnssI + \sin\rotvecnorm \Sskew[\rotvecaxis] + \br{1-\cos\rotvecnorm}\Sskew^2\br{\rotvecaxis},
\end{split}
\end{equation}
where $\exp\placeholder$ is the exponential map and $\Sskew$ is an operator that produces a skew-symmetric matrix such that $\Sskew[\tns{a}] \tns{b} = \tns{a} \times \tns{b} \ \forall \ \tns{a}, \tns{b} \in \R{3}$.
The calculation of the inverse of the Rodrigues formula is not straight forward.
For simplicity, it is abbreviated by the expression $\rotvec(\triad) = \rv(\triad)$ in the following.
In practice, Spurrier's algorithm \cite{Spurrier1978} can be used for the extraction of the rotation vector.
Let us consider two triads $\triad_1(\rotvec_{1})$ and $\triad_2(\rotvec_{2})$ with their respective rotation vectors~$\rotvec_{1}$ and~$\rotvec_{2}$.
They are related to each other by the relative rotation $\triad_{21}(\rotvec_{21})$.
The relative rotation is given by
\begin{equation}
\label{left_multi}
\begin{split}
\triad_{2}(\rotvec_{2}) &= \triad_{21}(\rotvec_{21}) \triad_1(\rotvec_{1})\\
& \Updownarrow\\
 \triad_{21}(\rotvec_{21}) &= \triad_2(\rotvec_{2}) \triad_1(\rotvec_{1})\tr,
\end{split}
\end{equation}
with the identity $\triad\tr=\triad\inv$ for all elements of $\SO$.
The relative rotation vector $\rotvec_{21}=\rv\br{\triad_{21}}$ describes the relative rotation between $\triad_1$ and $\triad_2$.
Rotation vectors are non-additive, \ie $\rotvec_{21} \ne \rotvec_{2} - \rotvec_{1}$.
For a more comprehensive treatment of this topic, the interested reader is referred to~\cite{Simo1986b, Cardona1988, Ibrahimbegovic1995, Romero2004, Meier2019, Betsch1998}.
In the following sections, both symbols~$\triad$ and~$\rotMat$ will be used to represent rotation tensors.

\subsection{Structure formulation}

We consider the interaction of beams with 2D surfaces of 3D continua (\btsso scenario) as well as 2D structural shell formulations (\btssh scenario).
In the following two subsections, we briefly outline the governing equations for both types of structures.
For the presented coupling formulations, the critical aspect is the mathematical description of the respective surface, which is identical in both cases.
The 2D surface domain in the reference configuration is~$\domainSurfaceref \subset \R3$, while~$\domainSurface \subset \R3$ is the domain in the current configuration.
Throughout this work, the subscript~$\placeholder_0$ indicates a quantity in the reference configuration.
The surface is parameterized with the two surface parameter coordinates $\xisurface \in \R{}$ and~$\etasurface \in \R{}$.
The current position~$\xsolid \in \R3$ of a material point relates to the reference position~$\Xsolid \in \R3$ via the structure displacement field $\usolid \in \R3$, \ie
\begin{equation}
	\label{eq:problem_formulation:solid_displacement}
	\xsolid\br{\xisurface, \etasurface} = \Xsolid\br{\xisurface, \etasurface} + \usolid\br{\xisurface, \etasurface}.
\end{equation}

\subsubsection{Solid formulation}

The solid is modeled as a 3D \boltz continuum, defined over the 3D domain $\domainSolidref \subset \R3$ in the reference configuration.
The boundary of the solid domain $\domainSolidSurfaceref$, represents the solid surface, \ie $\domainSolidSurfaceref = \domainSurfaceref$.
Accordingly,~$\domainSolid$ and~$\domainSolidSurface$ are the current solid domain and current solid surface, respectively.
The virtual work contributions of the solid domain read
\begin{equation}
\begin{split}
\dWsolid
=&
\intSolid{\Spk : \dE} \\
&-
\intSolid{\loadSolidBody \cdot \dusolid}
-
\intSolidNeumann{\loadSolidSurface \cdot \dusolid}
.
\end{split}
\end{equation}
Here, $\delta$ denotes the (total) variation of a quantity, $\Spk \in \R{3\times 3}$ is the second Piola--Kirchhoff stress tensor, $\E\in \R{3\times 3}$ is the work-conjugated Green--Lagrange strain tensor, $\loadSolidBody \in \R3$ is the body load vector and $\loadSolidSurface \in \R3$ are surface tractions on the Neumann boundary $\domainSolidNeumann \subset \domainSolidSurfaceref$.
Furthermore, $\F \in \R{3 \times 3}$ is the solid deformation gradient.
It should be noted that in the case of 3D solids, the kinematics of a material point, \ie $\Xsolid$, $\xsolid$ and $\usolid$ are not just defined on the solid surface but also inside of the solid volume.

\subsubsection{\kl shell formulation}

The shell formulation considered in this work is based on the \kl theory, which assumes that transverse shear deformations are negligible, making it suitable for thin shells.
Under the \kl assumption of straight and normal cross-sections, the shell continuum can solely be described by the midsurface.
The 2D shell domains in the reference and current configuration are $\domainSurfaceref$ and $\domainSurface$, respectively.
The total virtual work of the shell structure reads, \cf \cite{Kiendl2009,Belytschko2013,Bischoff2004},
\begin{equation}
\label{eq:weak_shell}
\dWsolid = \intSolidShellSurface{\shellStressTensor : \dshellStrain + \shellBendingMomentTensor : \dshellCurvature- \loadShellBody \dusolid}.
\end{equation}
Here, $\shellStressTensor$ is the stress resultant tensor, $\dshellStrain$ is the virtual membrane strain tensor, $\shellBendingMomentTensor$ is the bending moment tensor, $\dshellCurvature$ is the variation of the curvature tensor and $\loadShellBody$ is a body load (\ie surface load) acting on the shell.
For a more detailed description of the considered shell formulation and its finite element implementation, the interested reader is referred to~\cite{Kiendl2009}.

\subsection{\Gexact beam theory}
In this work the \gexact \sr beam theory is used to describe the embedded beams as 1D \cosserat continua, \eg \cite{Reissner1972, Meier2019, Simo1986a, Simo1986b}.
Each beam \cs along the beam centerline is described by six degrees of freedom, three positional and three rotational ones, thus resulting in six deformation modes of the beam: axial tension, bending (2$\times$), shear (2$\times$) and torsion.

The complete beam kinematics can be defined by a centerline curve $\rbeam(\sbeam) \in \R{3}$, connecting the
cross-section centroids, and a field of triads $\triadbeam(\sbeam) = \triadbeam(\rotvecbeam(\sbeam))$ defining the orientation of the beam cross-sections.
Here, $\sbeam \in [0, L] =: \domainBeamref$ is the arc-length coordinate along the beam centerline in the reference configuration and $L$ is the reference length of the beam.
The triad $\triadbeam$ is chosen such that the second and third basis vectors, $\gtriadbeam{2}$ and $\gtriadbeam{3}$, span the beam \cs, \ie the first triad basis vector $\gtriadbeam{1}$ is normal to the beam \cs.
A total hyperelastic stored-energy function of the \sr beam can be stated as
\begin{equation}
\Pbeam = \intBeamCenterline{ 
\tilde{\Pi}_{\nameinternal,\letterbeam}},
\end{equation}
with
\begin{equation}
\tilde{\Pi}_{\nameinternal,\letterbeam} =\frac{1}{2}(\srTension\tr \srTensionC \srTension
+
\srBending\tr \srBendingC \srBending).
\end{equation}
Here, $\srTension \in \R{3}$ is a material deformation measure representing axial tension and shear, $\srBending \in \R3$ is a material deformation measure representing torsion and bending, and $\srTensionC \in \R{3\times3}$ and $\srBendingC \in \R{3\times3}$ are \cs constitutive matrices.
The material force stress resultants $\tns{F}_B=\pfracinline{\tilde{\Pi}_{\nameinternal,\letterbeam}}{\srTension}$ and moment stress resultants $\tns{M}_B=\pfracinline{\tilde{\Pi}_{\nameinternal,\letterbeam}}{\srBending}$ can be derived from the hyperelastic stored-energy function.
Finally, the beam contributions to the weak form are given by
\begin{equation}
\dWbeam = \dPbeam + \dWbeamext,
\end{equation}
where $\dWbeamext$ is the virtual work of external forces and moments acting on the beam.
For a more comprehensive presentation of the weak form of the \gexact \sr beam theory, the interested reader is referred to~\cite{Meier2019}.

\subsection{\BtsscFull (\btssfull)}
\label{sec:problem_formulation:btssc}

The \btssfull method proposed in this work couples all six \cs degrees of freedom of the beam to the surface.
This is realized by coupling the positions of the beam centerline as well as the orientation of the beam \cs to the surface.
One advantage of a 1D-2D coupling approach solely enforced at the beam centerline is the decoupling of the positional and rotational coupling conditions, \ie both of them can be formulated independently.
For embedded 1D geometrically exact beams in 3D volumes such an approach has recently been presented in~\cite{Steinbrecher2022}.
The same general strategy is followed here for \btss coupling problems, where we define two sets of coupling constraints, the positional coupling constraints (\btsstrans) and the rotational coupling constraints (\btssrot).
With this split, the total \btssfull coupling contribution to the weak form reads,
\begin{equation}
\dcouplingPotentialMortar =
\dcouplingPotentialPosMortar + \dcouplingPotentialRotMortar,
\end{equation}
where $\dcouplingPotentialPosMortar$ and $\dcouplingPotentialRotMortar$ are the virtual work contributions from the positional and rotational coupling conditions, respectively.

\subsubsection{Closest point projection}
\begin{figure*}
\centering
\subfigure[]{\label{fig:problem:btss_possible:curved_wo_offset}\includegraphicsdpi{300}{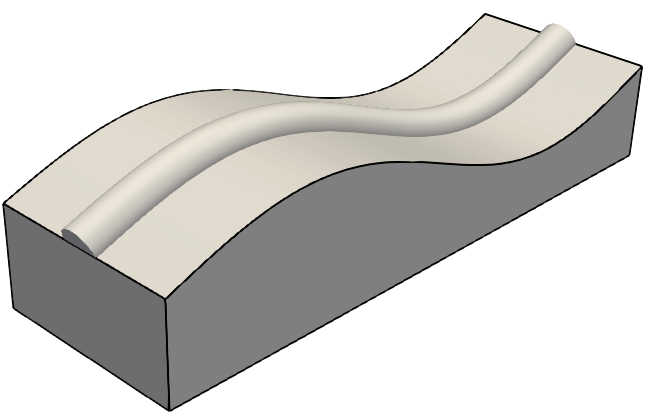}}
\hfill
\subfigure[]{\label{fig:problem:btss_possible:curved}\includegraphicsdpi{300}{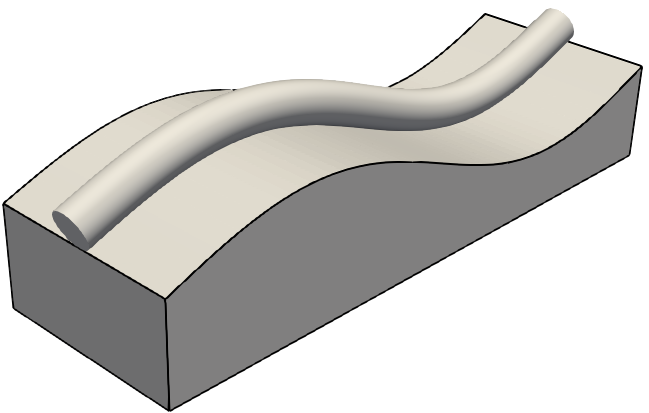}}
\hfill
\subfigure[]{\label{fig:problem:btss_possible:flipped}\includegraphicsdpi{300}{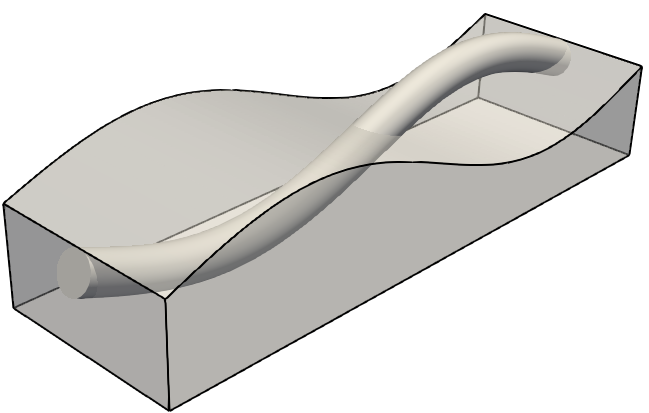}}
\subfigure[]{\label{fig:problem:btss_possible:curved_wo_offset_shell}\includegraphicsdpi{300}{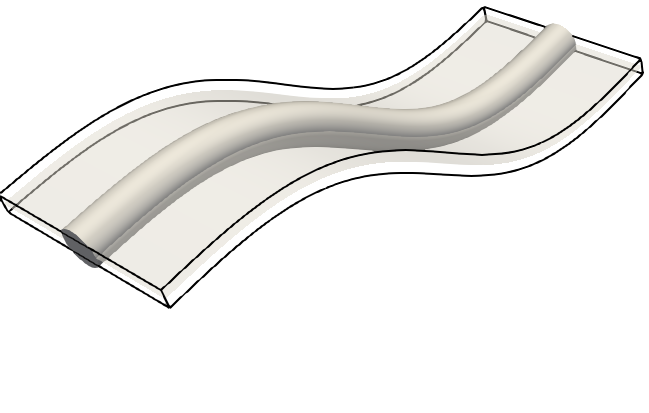}}
\hfill
\subfigure[]{\label{fig:problem:btss_possible:curved_shell}\includegraphicsdpi{300}{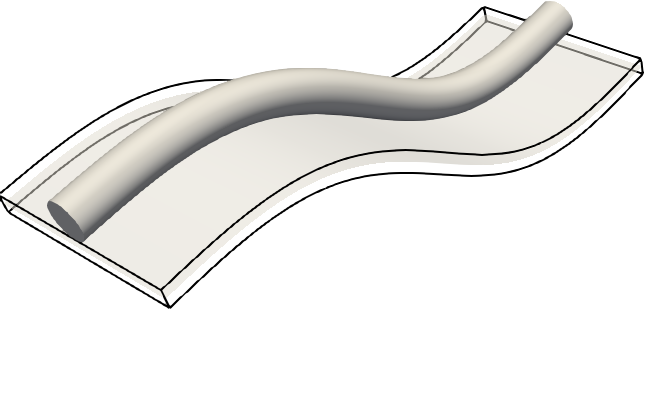}}
\hfill
\subfigure[]{\label{fig:problem:btss_possible:flipped_shell}\includegraphicsdpi{300}{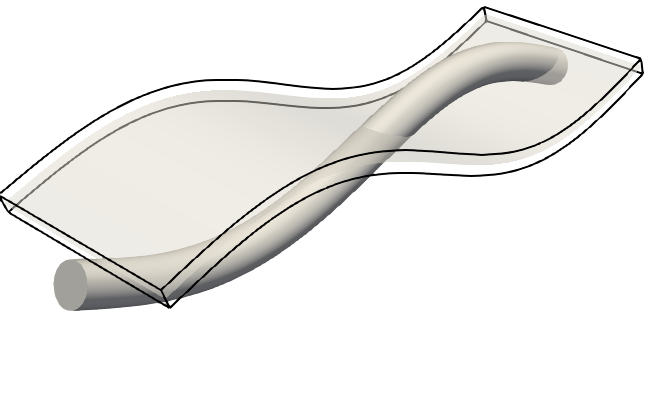}}
\caption{%
Illustration of possible \btss coupling problems.
For the \btsso scenario: \subref{fig:problem:btss_possible:curved_wo_offset} beam centerline on a surface, \subref{fig:problem:btss_possible:curved}
beam centerline offset by the \cs radius in surface normal direction, and \subref{fig:problem:btss_possible:flipped} a general non-matching case.
For the \btssh scenario: \subref{fig:problem:btss_possible:curved_wo_offset_shell} beam centerline on a shell midsurface, \subref{fig:problem:btss_possible:curved_shell}
beam centerline offset by the \cs radius and half of the shell thickness in shell midsurface normal direction, and \subref{fig:problem:btss_possible:flipped_shell} a general non-matching case.
}
\label{fig:problem:btss_possible}
\end{figure*}
In the considered \btssfull coupling problem, \cf Figure~\ref{fig:problem_formulation:btssc_problem}, no requirements on the initial beam position relative to the surface exist.
This is illustrated for both considered scenarios, \btsso and \btssh, in \Cref{fig:problem:btss_possible}.
The coupling scheme must be applicable to cases where the beam centerline curve lies on the surface (\cf \Cref{fig:problem:btss_possible:curved_wo_offset,fig:problem:btss_possible:curved_wo_offset_shell}) and to cases where the beam centerline is offset by a physically motivated distance in the surface normal direction (\cf \Cref{fig:problem:btss_possible:curved,fig:problem:btss_possible:curved_shell}).
Additionally, the presented coupling schemes also account for general cases without strict requirements on the reference placement of the beam centerline relative to the surface (\cf \Cref{fig:problem:btss_possible:flipped,fig:problem:btss_possible:flipped_shell}).
The only requirement considered in this work is a unique closest point projection of each beam centerline point onto the surface.
For the envisioned application cases, it can be assumed that a unique solution of the closest point projection exists in the vicinity of each beam centerline point $\rbeamO$, \cf \cite{Konyukhov2008}.
In the reference configuration each point $\rbeamO(\xibeampaper)$ on the beam centerline is assigned to a corresponding closest point $\Xsolid(\xisurfaceCpp, \etasurfaceCpp)$ on the surface, where $\xisurfaceCpp = \xisurfaceCpp(\xibeampaper)$ and $\etasurfaceCpp = \etasurfaceCpp(\xibeampaper)$ are the surface parameter coordinates of the closest point.
The closest point can be found by formulating a unilateral minimal distance problem in the reference configuration:
\begin{equation}
\label{eq:problem_formulation:minimal_distance_problem}
\surfaceNormalDistanceULO(\xibeampaper)
:=
\min_{\xisurface, \etasurface} \surfaceNormalDistance(\xibeampaper, \xisurface, \etasurface)
=
\surfaceNormalDistance(\xibeampaper, \xisurfaceCpp, \etasurfaceCpp)
\end{equation}
with
\begin{equation}
\surfaceNormalDistance(\xibeampaper, \xisurface, \etasurface) =  \norm{\rbeamO(\xibeampaper) - \Xsolid(\xisurface, \etasurface)}.
\end{equation}
The two orthogonality conditions obtained from the minimal distance problem \eqref{eq:problem_formulation:minimal_distance_problem} read
\begin{equation}
\label{eq:problem_formulation:orthogonality_conditions}
\begin{split}
\dXsoliddxi(\xisurface, \etasurface)\tr \br{\rbeamO(\xibeampaper) - \Xsolid(\xisurface, \etasurface)} = 0,
\\
\dXsoliddeta(\xisurface, \etasurface)\tr \br{\rbeamO(\xibeampaper) - \Xsolid(\xisurface, \etasurface)} = 0.
\end{split}
\end{equation}
For a given beam coordinate $\xibeampaper$, these conditions can be solved for the unknown surface coordinates $\xisurface$ and $\etasurface$.
The non-trivial solution of \eqref{eq:problem_formulation:orthogonality_conditions} requires the surface directors~$\dXsoliddxi = \pfracinline{\Xsolid}{\xisurface}$ and $\dXsoliddeta = \pfracinline{\Xsolid}{\etasurface}$ to be orthogonal to the relative vector between the surface point and the beam centerline point, \ie this relative vector is parallel to the outward pointing surface normal vector $\surfaceNormalO \in \R{3}$,
\begin{equation}
\label{eq:problem_formulation:position_coupling_reference}
\rbeamO(\xibeampaper) - \Xsolid(\xisurfaceCpp, \etasurfaceCpp) 
= \surfaceNormalDistanceULO(\xibeampaper) \surfaceNormalO(\xisurfaceCpp, \etasurfaceCpp)
\end{equation}
with
\begin{equation}
\label{eq:problem_formulation:normal_vector_ref}
\surfaceNormalO(\xisurface, \etasurface) = \normalize{\dXsoliddxi(\xisurface, \etasurface) \times \dXsoliddeta(\xisurface, \etasurface)}.
\end{equation}

\subsection{Positional \btssc (\btsstrans)}
\label{sec:variants}

\begin{figure}
\centering
\subfigure[]{\label{fig:problem:trans:cons}\includegraphics[page=1,scale=1]{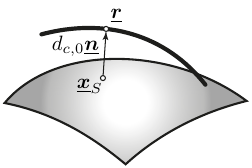}}
\hfill
\subfigure[]{\label{fig:problem:trans:ref}\includegraphics[page=3,scale=1]{figures/problem_formulation_trans_variants.pdf}}
\hfill
\subfigure[]{\label{fig:problem:trans:disp}\includegraphics[page=2,scale=1]{figures/problem_formulation_trans_variants.pdf}}
\caption{%
Illustration of the different positional \btssLong (\btsstrans) coupling variants.
\subref{fig:problem:trans:cons} Consistent positional coupling (\nameConsistent) via the surface normal vector, \subref{fig:problem:trans:ref} forced reference configuration coupling (\nameReference) by forcing beam centerline points to lie on the surface and \subref{fig:problem:trans:disp} displacement coupling (\nameDisplacement), where the displacement of beam centerline and surface are coupled.
The surface-to-surface equivalents of the \nameReference and \nameDisplacement variants are commonly used in classical \sutsu mesh tying problems~\cite{Puso2004}.
}
\label{fig:problem:trans}
\end{figure}

In this section, three different variants of the \btsstrans coupling constraints are presented. They will be compared with each other in more detail in \Cref{sec:examples}.
The first presented variant in \Cref{sec:variants_consisent} is consistent with the kinematic relations between beam centerline and surface.
The resulting coupling terms contain the surface normal vector, \ie the coupling terms become non-linear.
Furthermore, the second derivative of the surface normal vector is required for a consistent linearization
of the problem in tangent-based nonlinear solution schemes (such as the \nr algorithm).
To avoid this computationally expensive linearization in each iteration of the nonlinear solution scheme, two additional variants to formulate the positional coupling constraints will be investigated, \cf\Cref{sec:variants_ref,sec:variants_disp}.
These variants are also commonly used in classical \sutsu mesh tying problems \cite{Puso2004}.
Both of them do not require an evaluation of the current surface normal vector or its derivatives, and the resulting coupling operators only depend on the reference configuration, \ie they are constant.
The different coupling variants are visualized in \Cref{fig:problem:trans}.

\subsubsection{Consistent positional coupling (\nameConsistent)}
\label{sec:variants_consisent}

The \btsstrans coupling constraints are exclusively formulated along the beam centerline and couple the beam and solid material points associated by \eqref{eq:problem_formulation:position_coupling_reference} to each other.
For the considered consistent variant, the surface normal distance $\surfaceNormalDistanceUL$ at each beam centerline point shall be constant over the simulation (pseudo-)time, \ie $\surfaceNormalDistanceUL \equiv \surfaceNormalDistanceULO$.
Therefore, the coupling equations in the current configuration can be formulated as
\begin{equation}
\label{eq:problem_formulation:position_coupling}
\rbeam(\xibeampaper) - \xsolid(\xisurfaceCpp, \etasurfaceCpp) - \surfaceNormalDistanceULO(\xibeampaper) \, \surfaceNormal(\xisurfaceCpp, \etasurfaceCpp) = \tnsO \quad \text{on} \quad \domainCoupling.
\end{equation}
The current normal vector is defined in analogy to the reference normal vector \eqref{eq:problem_formulation:normal_vector_ref}, \ie
\begin{equation}
\label{eq:problem_formulation:normal_vector_cur}
\surfaceNormal(\xisurfaceCpp, \etasurfaceCpp) = \normalize{\dxsoliddxi(\xisurfaceCpp, \etasurfaceCpp) \times \dxsoliddeta(\xisurfaceCpp, \etasurfaceCpp)},
\end{equation}
with the current surface directors $\dxsoliddxi = \pfracinline{\xsolid}{\xisurface}$ and $\dxsoliddeta = \pfracinline{\xsolid}{\etasurface}$.
The constraints \eqref{eq:problem_formulation:position_coupling} are enforced along the one-dimensional coupling domain $\domainCoupling \subseteq \domainBeam[,0]$ between the beam centerline and the surface, \ie the part of the beam that is coupled to the surface.
In the following considerations, the explicit dependency on the beam and solid parameter coordinates will mostly be omitted for improved readability.

In the remainder of this work, the positional coupling constraints \eqref{eq:problem_formulation:position_coupling} will be referred to as the \emph{consistent} (\nameConsistent) surface coupling variant.
The name refers to the fact that the coupling definition is consistent with the kinematic relations between surface and beam centerline, \cf \Cref{fig:problem:trans:cons}.
Furthermore, it will be shown that this variant leads to vanishing constraint forces in the (undeformed) reference configuration and exact conservation of linear and angular momentum in the discretized coupled system, \cf \Cref{sec:discretization:conservation_laws}.

The Lagrange multiplier method is used to weakly enforce the coupling constraints \eqref{eq:problem_formulation:position_coupling}.
Therefore, a Lagrange multiplier vector field $\lagrangePos(\sbeam) \in \R{3}$, defined along the beam centerline, is introduced.
The total Lagrange multiplier potential reads:
\begin{equation}
\couplingPotentialPosMortar = \intCoupling{\lagrangePos\tr \br{\rbeam - \xsolid - \surfaceNormalDistanceULO \surfaceNormal}}.
\end{equation}
Variation of the Lagrange multiplier potential leads to the constraint contribution to the weak form,
\begin{equation}
\label{eq:problem_formulation:trans_weak_form}
\begin{split}
\dcouplingPotentialPosMortar = &
\underbrace{\intCoupling{\dlagrangePos\tr \br{\rbeam - \xsolid - \surfaceNormalDistanceULO \surfaceNormal}}}_{\dWposLambda}
\\ &+
\underbrace{\intCoupling{\lagrangePos\tr \br{\drbeam - \dxsolid - \surfaceNormalDistanceULO \dsurfaceNormal} }}_{-\dWposC}.
\end{split}
\end{equation}
Therein, $\dWposLambda$ and $\dWposC$ are the variational form of the coupling constraints and the virtual work of the Lagrange multiplier field $\lagrangePos$, respectively.
It is well-known from geometrically exact beam theory that the variation of the centerline position $\drbeam$ is work-conjugated with the infinitesimal forces acting on the beam centerline, \ie $\lagrangePos \mathrm d \sbeam$.
Therefore, the Lagrange multiplier field $\lagrangePos$ can be directly interpreted as the coupling line load acting on the beam centerline.
On the solid side, the variation of the solid displacement $\dxsolid$ is work conjugated with the infinitesimal forces $-\lagrangePos \mathrm d \sbeam$ acting on the solid, \ie the (negative) Lagrange multiplier field also acts as a line load on the solid.
Additionally, the term~$\surfaceNormalDistanceULO \lagrangePos\tr \dsurfaceNormal \mathrm d \sbeam$ arises, which represents an infinitesimal moment contribution of the coupling line load on the solid.
The drawback of this variant is that the weak form contains the surface normal vector variation.
This has two important implications on the presented mixed-dimensional coupling approach: (i)~The positional coupling operators become non-linear, \ie they depend on the current configuration and have to be re-evaluated for each iteration of the nonlinear solution scheme.
(ii)~Second derivatives of the surface normal vector have to be computed for a consistent linearization of $\dsurfaceNormal$ as required for tangent-based nonlinear solution schemes (such as \nr iterations).

\subsubsection{Forced reference configuration coupling (\nameReference)}
\label{sec:variants_ref}

The considered 1D-2D line-to-surface coupling constraints are very similar to the ones in classical 2D-2D \sutsu coupling problems, \cf \cite{Puso2004,Park2002,Dohrmann2000}.
In such problems, the space continuous interfaces are usually matching, \ie the normal distance vanishes.
Even if the surfaces do not match exactly, \eg due to incompatible CAD files, the influence of the surface normal vector can usually be neglected since it is in the range of the discretization error.
Therefore, the coupling constraints \eqref{eq:problem_formulation:position_coupling} can be simplified to
\begin{equation}
\label{eq:problem_formulation:position_coupling_ref}
\rbeam - \xsolid = \tnsO \quad \text{on} \quad \domainCoupling.
\end{equation}
This type of positional coupling constraint will be referred to as the \emph{forced reference configuration} surface coupling (\nameReference).
The Lagrange multiplier coupling contributions to the global weak form read
\begin{align}
\dWposLambdaSurfRef &= \intCoupling{\dlagrangePos\tr \br{\rbeam - \xsolid}}
\\
-\dWposCSurfRef &= \intCoupling{\lagrangePos\tr \br{\drbeam - \dxsolid} }.
\end{align}
In this case, the surface normal vector is not contained in the resulting 
coupling equations, thus simplifying the numerical evaluation of the coupling terms.
However, the coupling constraints \eqref{eq:problem_formulation:position_coupling_ref} in the reference configuration are only fulfilled if the beam centerline lies exactly on the surface, \ie~$\surfaceNormalDistanceULO \equiv 0$.
If the beam centerline is not a subset of the surface, the coupling constraints \eqref{eq:problem_formulation:position_coupling_ref} will lead to non-vanishing virtual work contributions in the reference configuration, \ie initial stresses and deformations in the unloaded coupled system.
In other words, the \nameReference coupling conditions force the beam centerline to exactly lie on the surface, which is illustrated in \Cref{fig:problem:trans:ref}.

\subsubsection{Displacement coupling (\nameDisplacement)}
\label{sec:variants_disp}

Another alternative coupling approach in \sutsu mesh tying is to directly couple the displacements instead of the positions in \eqref{eq:problem_formulation:position_coupling_ref}, \cf \cite{Puso2004}.
This variant will be referred to as the \emph{displacement} surface coupling (\nameDisplacement).
The \nameDisplacement coupling constraints read,
\begin{equation}
\label{eq:problem_formulation:position_coupling_disp}
\ubeam - \usolid = \tnsO \quad \text{on} \quad \domainCoupling,
\end{equation}
with the beam centerline displacement $\ubeam = \rbeam - \rbeamO$.
The Lagrange multiplier coupling contributions to the global weak form are
\begin{align}
\dWposLambdaSurfDisp &= \intCoupling{\dlagrangePos\tr \br{\ubeam - \usolid}}
\\
\label{eq:problem_formulation:position_coupling_disp_virtual_work}
-\dWposCSurfDisp &= \intCoupling{\lagrangePos\tr \br{\dubeam - \dusolid} }.
\end{align}
As is the case for the \nameReference variant, the normal vector does not appear in the coupling constraints.
In this case, the coupling conditions are always fulfilled in the reference configuration no matter if the initial geometries of beam centerline and surface are matching or not.
In \cite{Puso2004} it is demonstrated that displacement coupling \eqref{eq:problem_formulation:position_coupling_disp} can lead to a coupling formulation that does not conserve angular momentum.
This can be shown by inserting a constant virtual rotation $\virtRot$, \ie $\dubeam = \virtRot \times \rbeam$ and $\dusolid = \virtRot \times \xsolid$, into~\eqref{eq:problem_formulation:position_coupling_disp_virtual_work}.
To guarantee conservation of angular momentum the resulting virtual work has to vanish, \cf \cite{Puso2004}.
This gives the condition for conservation of angular momentum
\begin{equation}
\intCoupling{ \br{\virtRot \times \br{\rbeam - \xsolid}}\tr \lagrangePos } = 0.
\end{equation}
This condition is only fulfilled if $\rbeam = \xsolid$, \ie for matching interfaces.
For general configurations of the beam and the solid, \ie when the beam centerline is offset in surface normal direction, conservation of angular momentum is violated by the \nameDisplacement variant.
This can also be interpreted from a mechanical point of view: displacement coupling of two points (a point on the beam centerline and the corresponding projection point on the surface) that do not coincide in the reference configuration, \cf \Cref{fig:problem:trans:disp}, leads to a non-physical coupling moment, which violates the conservation of angular momentum.

\subsection{Rotational \btssc (\btssrot)}

Rotational 1D-3D beam-to-volume (\btsvr) coupling between an embedded \gexact beam with a \boltz continuum has recently been presented and thoroughly discussed~\cite{Steinbrecher2022}.
There, it has been shown that constraining the relative rotation (pseudo-)vector $\rotvecbeamsolid$ between the current beam triad $\triadbeam$ and a suitable volume triad $\triadVolume$ along the beam centerline leads to an objective coupling scheme.
This approach is in accordance to general \cs interaction laws within the \gexact beam theory, \cf \cite{Meier2023}.
This type of rotational coupling scheme can also be adopted for the presented case of 1D-2D \btss coupling problems.
The general approach is the same as in \cite{Steinbrecher2022}, but instead of a volume triad field, a suitable \emph{surface} triad field has to be constructed.
This construction procedure will be presented in \Cref{sec:surface_triad}.
The rotational coupling constants constrain the relative rotation vector between the beam triad and a corresponding surface triad $\triadsurface$, \ie
\begin{equation}
\label{eq:problem_formulation:rotation_coupling}
\rotvecbeamsolid = \tnsO \quad \text{on} \quad \domainCoupling,
\end{equation}
with
\begin{equation}
\label{eq:problem_formulation:rotation_constraint_equations}
\rotvecbeamsolid = \rv(\triadsurface \triadbeam\tr).
\end{equation}
The Lagrange multiplier method is used to weakly enforce the rotational coupling constraints \eqref{eq:problem_formulation:rotation_coupling}.
The corresponding weak form has been derived and thoroughly discussed in \cite[Section 4.3.2]{Steinbrecher2022} and will not be stated here.

\section{Surface triad field}
\label{sec:surface_triad}
\newcommand{\triadsurfaceAverage}{\triad_{\nameSurf AVERAGE}}

The rotational coupling conditions \eqref{eq:problem_formulation:rotation_coupling} constrain the relative rotation vector $\rotvecbeamsolid$ between the beam \cs triad~$\triadbeam$ and a corresponding \emph{surface} triad $\triadsurface$.
In the \btsso scenario, the surface is part of a 3D \boltz continuum, which inherently does not possess any rotational degrees of freedom.
In the \btssh scenario, we consider a \kl shell formulation.
Under this formulation, the shell midsurface is solely defined by its positional field, and the surface normal director is the kinematic normal to the midsurface.
As a result, the \kl shell formulation also does not introduce any rotational degrees of freedom.
Thus, for both considered scenarios, a suitable surface triad field must be constructed as a function of the surface position field.
The construction of triad fields inside solid volumes has very recently been thoroughly discussed and analyzed in~\cite{Steinbrecher2022}.
There, two important attributes of the constructed triad field are identified: (i) The triad field has to be invariant with respect to an arbitrary rigid body rotation, such that the rotational coupling constraints \eqref{eq:problem_formulation:rotation_coupling} lead to an objective discrete coupling formulation.
(ii) The resulting triad field should not constrain shear deformations in the beam \cs plane, as this can result in spurious stiffening / locking effects of the coupled system.
In this section, we present a novel surface triad construction for \btss coupling problems, that fulfills both aforementioned properties.
The surface triad is constructed directly based on the surface kinematics, \ie the two surface basis vectors and the surface normal vector.
In theory, when 2D surfaces of 3D volumes are considered (\btsso scenario), the solid volume triad definitions from \cite{Steinbrecher2022} could also be employed.
However, there are significant drawbacks to this approach, which are outlined in detail in \Cref{sec:appendix_solid_surface_triad}.

\begin{figure*}
\centering
\includegraphics[scale=1]{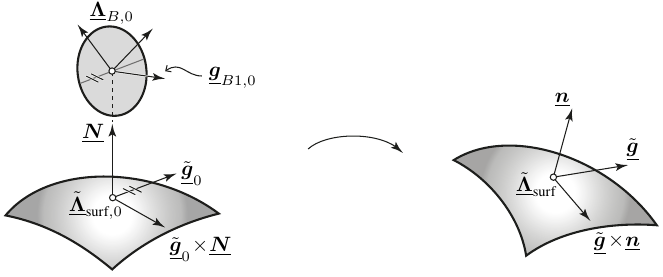}
\caption{Construction of the surface triad.}
\label{fig:surface_triad:surface_triad_construction}
\end{figure*}

The proposed construction of the surface triad is based on a material director $\gtriadtilde{}$ lying on the surface, in combination with the surface normal vector.
The obvious and intuitive choice for this surface material director is the intersection between the beam \cs plane and the surface tangent plane in the reference configuration, \cf \Cref{fig:surface_triad:surface_triad_construction}, which reads
\begin{equation}
\gtriadtilde{0} = \normalize{\surfaceNormalO \times \gtriadbeamO{1}}.
\end{equation}
Theoretically, this definition of the surface material director can result in a singularity if the beam \cs and the surface tangent plane are parallel to each other.
However, since this would mean that the beam centerline is normal to the surface, this singularity will not be relevant for practical applications.
The surface triad in the reference configuration can subsequently be constructed based on the surface material director and the surface normal vector, \ie
\begin{equation}
\label{eq:triad:reference_triad}
\triadsurfaceRel = \matrix{
	\gtriadtilde{0},
	\surfaceNormalO,
	\gtriadtilde{0} \times \surfaceNormalO
}.
\end{equation}
The normalized surface material director in the current configuration $\gtriadtilde{}$ is calculated by applying the surface deformation gradient $\Fsurface$ to the material director $\gtriadtilde{0}$ in the reference configuration, \ie $\gtriadtilde{} =	\Fsurface \, \gtriadtilde{0} / \| \Fsurface \, \gtriadtilde{0} \|$.
With the current surface material director, the surface triad in the current configuration can be constructed in analogy to~\eqref{eq:triad:reference_triad}, \ie
\begin{equation}
\label{eq:triad:current_triad}
\triadsurfaceRef = \matrix{
	\gtriadtilde{},
	\surfaceNormal,
	\gtriadtilde{} \times \surfaceNormal
}.
\end{equation}
In a final step, the actual surface triad used for evaluation of the coupling terms has to be offset by a constant rotation, such that the rotational constraint equations \eqref{eq:problem_formulation:rotation_constraint_equations} are fulfilled in the reference configuration.
The final surface triad reads,
\begin{equation}
\label{eq:triad:surface_triad}
\triadsurface = \triadsurfaceRef \triadsurfaceRel\tr \triadbeamO.
\end{equation}
With this definition, it is straight-forward to show that the surface triad in the reference configuration is equal to the beam reference triad, \ie $\triadsurfaceO = \triadbeamO$ and therefore, the rotational coupling constraints are fulfilled in the reference configuration.
It can also be shown that the surface triad definition \eqref{eq:triad:surface_triad} is invariant with respect to a superposed rigid body rotation, thus fulfilling requirement (i) stated above.
Furthermore, since the surface triad is constructed based on a single material director $\gtriadtilde{}$ and the surface normal vector, a constraining of shear deformations on the surface can not occur.
Therefore, the presented surface triad also fulfills requirement (ii).

\section{Spatial discretization}
\label{sec:discretization}

In this work, the spatial discretization is exclusively based on the finite element method.
In the following, a subscript $\placeholder_h$ refers to an interpolated field quantity.

Two types of structures (solid and shell) are considered in this work.
In both cases, an isoparametric Bubnov--Galerkin discretization approach is employed, \ie the position, displacement, and virtual displacement fields are discretized with the same shape functions.
For 3D solid bodies, this discretization is exclusively based on standard $\C{0}$-continuous Lagrangian finite elements.
However, in the case of \kl shells, this approach is insufficient, as the weak form \eqref{eq:weak_shell} explicitly contains the first and second derivatives of the surface displacement field.
Consequently, the approximation of the shell surface must be at least $G^1$-continuous.
To meet this requirement, isogeometric analysis (IGA) is employed for the discretization of the \kl shell, as \nurbs-based basis functions enable the construction of approximations with arbitrary orders of continuity.
For further details, the interested reader is referred to~\cite{Kiendl2009}.

In the following derivation of the discretized coupling terms, only the discretized surface field is required, which is parameterized by the two surface parameter coordinates $\xisurface$ and $\etasurface$.
The spatial interpolation of the surface is given by
\begin{align}
\Xsolidh &= \sumsurface{\Nsolidn\br{\xisurface, \etasurface} \qXsolidn}
\\
\usolidh &= \sumsurface{\Nsolidn\br{\xisurface, \etasurface} \qsolidn}
\\
\dusolidh &= \sumsurface{\Nsolidn\br{\xisurface, \etasurface} \dqsolidn}.
\end{align}
Here, $\nsurface$ represents the number of surface nodes (FEM) or surface control points (IGA), and $\Nsolidn \in \R{}$ is the corresponding shape function, which can represent finite element basis functions or \nurbs basis functions depending on the discretization approach.
Furthermore, $\qXsolidn \in \R3$, $\qsolidn \in \R3$, and~$\dqsolidn \in \R3$ denote the reference position, displacement, and virtual displacement associated with the node or control point $\indexsolid$, respectively. 

The beam centerline interpolation considered in this work is exclusively based on third-order Hermitian polynomials, \cf \cite{Meier2019, Vetyukov2014}.
In this case, each node contains six centerline degrees of freedom, the nodal position $\qxbeamer \in \R{3}$ and the nodal centerline tangent $\qxbeamet \in \R{3}$ at the beam node $\indexbeam$.
This yields a~$\C{1}$-continuous beam centerline interpolation according to
\begin{equation}
\label{eq:discret_fe_beam_centerline}
\rbeamh = \sumbeam{\Nbeamr(\xibeam) \qxbeamer + \Nbeamt(\xibeam) \qxbeamet}.
\end{equation}
Here, $\nbeam$ is the number of beam nodes and $\Nbeamr \in \R{}$ and~$\Nbeamt \in \R{}$ denote the Hermite shape functions for the positional and tangential degrees of freedom of beam node $\indexbeam$.
Both shape functions have the scalar beam centerline parameter coordinate $\xibeam$ as argument.
At this point it is important to emphasize that the positional Hermite shape functions fulfill the partition of unity property, \ie $\sumbeam{\Nbeamr} \equiv 1$, \cf \cite{Meier2014}.
To improve readability the beam centerline interpolation \eqref{eq:discret_fe_beam_centerline} is rewritten in the following way
\begin{equation}
\rbeamh = \sumbeam{\Nbeamn\br{\xibeam}} \qxbeamn,
\end{equation}
with
\begin{align}
\Nbeamn &= \matrix{\Nbeamr \matI^{3 \times 3} & \Nbeamt \matI^{3 \times 3}} \in \R{3\times 6}
\\
\qxbeamn &= \matrix{\qxbeamer \\ \qxbeamet} \in \R{6}.
\end{align}
Here, $\Nbeamn$ is the matrix with the node-wise assembled beam centerline shape functions and $\qxbeamn$ is the corresponding generalized nodal position vector.
The discretized variation of the beam centerline position is given by
\begin{equation}
\drbeamh = \sumbeam{\Nbeamn} \dqbeamn,
\qquad \text{with} \qquad
\dqbeamn = \matrix{\dqbeamer \\ \dqbeamet}.
\end{equation}
Here $\dqbeamer \in \R{3}$ and $\dqbeamet \in \R{3}$ are the variations of the discrete nodal position and tangent, respectively.
An objective and path-independent spatial interpolation of the beam \cs rotations is a non-trivial task.
The rotational interpolation only affects the \btssrot coupling terms, which are based on the \btsvr method derived in \cite{Steinbrecher2022}, and will therefore not be stated here.
For a more detailed discussion on this topic the interested reader is referred to \cite{Meier2019,Steinbrecher2022}.

\subsection{Evaluation of surface normal field}
\label{sec:discretization:surface_normal_field}

The closest point projection \eqref{eq:problem_formulation:minimal_distance_problem} of a point along the beam centerline to the surface requires a $\C{0}$-continuous normal field to guarantee a unique solution.
If the structure discretization is based on isogeometric elements with higher order continuity, then the surface normal field can be directly calculated from the kinematic description of the discretized surface.
The resulting surface normal field is at least $\C{0}$-continuous and a unique closest point projection can be guaranteed.
If a standard $\C{0}$-continuous Lagrangian finite element interpolation is employed, the surface normal field obtained from the kinematic description of the discretized surface is not continuous.
This can result in an undefined closest point projection.
However, the \btssc schemes presented in this work is also applicable to such discretizations.
This is achieved by constructing a $\C{0}$-continuous normal field based on averaged nodal normal vectors, as is common in \sutsu interaction problems, \cf \cite{Yang2005, Popp2009}.

\begin{figure*}
\centering
\subfigure[\label{fig:discretization:averaged_normals_discontinuous}]{\includegraphics[scale=1]{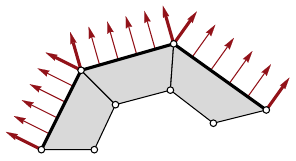}}
\hfill
\subfigure[\label{fig:discretization:averaged_normals_averaged}]{\includegraphics[scale=1]{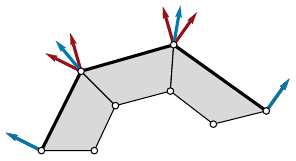}}
\hfill
\subfigure[\label{fig:discretization:averaged_normals_averaged_interpolated}]{\includegraphics[scale=1]{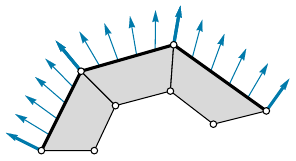}}
\caption{%
Illustration of the constructed $\C{0}$-continuous surface normal field, for an exemplary planar problem with three $\C{0}$-continuous Lagrangian finite elements.
\subref{fig:discretization:averaged_normals_discontinuous} Discontinuous standard surface normal field based on the finite element surface kinematics, \subref{fig:discretization:averaged_normals_averaged} averaged nodal normals and \subref{fig:discretization:averaged_normals_averaged_interpolated} $\C{0}$-continuous interpolated averaged nodal normal field.
}
\label{fig:discretization:averaged_normals}
\end{figure*}

The main idea behind the construction of an averaged surface normal field is illustrated in \Cref{fig:discretization:averaged_normals}.
An averaged nodal normal is defined at each surface node $\nodesurface$ as
\begin{equation}
\nodeNormalAverage = \normalize{
\sum_{\esolidName=1}^{\nodeNormalNumber} {\nodeNormal}
},
\end{equation}
where $\nodeNormal$ is the outward pointing surface normal vector of element $\esolidName$, evaluated at node $\nodesurface$.
Furthermore, $\nodeNormalNumber$ represents the number of adjacent facets at node $\nodesurface$.
The final normal vector field is then defined via a FE interpolation, \ie
\begin{equation}
\surfaceNormalh(\xisurface, \etasurface) = \normalize{ \sumsurface{ N_\nodesurface(\xisurface, \etasurface) \nodeNormalAverage }}.
\end{equation}
Such a surface normal field is guaranteed to be $\C{0}$-continuous, \ie it mimics a $\C{1}$-continuous surface interpolation.
However, one should admit that this procedure increases the computational effort required to evaluate the normal field and its derivatives.
Additionally, the connectivity between element degrees of freedom is increased, as the normal on a solid face element depends on the degrees of freedom of the adjacent facets.

\begin{remark}
Due to the averaging procedure, the resulting averaged normal is not point-wise orthogonal to the solid surface, \cf \Cref{fig:discretization:averaged_normals}.
With the definition of the surface triad \eqref{eq:triad:current_triad}, this would result in a non-orthonormal tensor $\triadsurface \notin \SO$.
Therefore, the actual point-wise orthogonal normal vector on the surface, not the averaged normal vector, is used in the evaluation of the surface triad in the \btsso scenario.
\end{remark}

\subsection{Uncoupled problem}
Inserting the finite element approximations into the weak form of the equilibrium equations for the beam and structure (solid or shell) problem and applying a standard \nr procedure as nonlinear solution scheme yields the linearized global system of equations for the uncoupled problem,
\begin{equation}
\label{eq:discretization:system_uncoupled}
\matrix{
\Kss & \matO & \matO \\
\matO & \Krr & \Krt \\
\matO & \Ktr & \Ktt
}
\vector{\DQsolid \\ \DQbeam \\ \DQbeamrot}
=
\begin{bmatrix}
-\Rsolid \\
-\Rbeam \\
-\RbeamRot
\end{bmatrix}.
\end{equation}
Therein, $\Kss$ is the structure tangent stiffness matrix, $\DQsolid$ is the increment of the discrete structure degrees of freedom and $\Rsolid$ is the residual force vector associated with the structure degrees of freedom.
The beam degrees of freedom are split up into positional and rotational degrees of freedom, indicated by the subscripts $\letterbeamcenterline$ and $\letterbeamrot$, respectively.
Accordingly, $\KbeamPlace$ are the beam tangent stiffness matrices, $\DQbeamPlace$ are the increments of the beam degrees of freedom, and $\RbeamPlace$ are the residual force vectors associated with the respective beam degrees of freedom.

The global structure displacement vector $\qsolid$ might also contain degrees of freedom not related to the coupling surface, \ie the number of total structure degrees of freedom can be larger than the number of surface degrees of freedom.
However, a split of the structure degrees of freedom into coupling surface and other degrees of freedom is not introduced in what follows for improved readability.

\begin{remark}
	\label{rem:rotation_updates}
	The beam formulation employed in this work parametrizes the total orientation of beam node $\indexbeam$ with the rotation (pseudo-)vector $\qxbeamrot$.
	It should be pointed out that the global vector $\DQbeamrot$ in \eqref{eq:discretization:system_uncoupled} represents the nodal \emph{multiplicative} rotational updates $\qxbeamrotincrement$.
	The update of the nodal orientations from non-linear iteration $i$ to $i+1$ is performed by $\qxbeamrotnp = \rv(\triad(\qxbeamrotincrement) \triad(\qxbeamrotn))$, \cf {\Cref{sec:finite_rotations}}.
\end{remark}

\subsection{Mortar-type coupling of beam-to-surface normal distance (\btsstrans)}
\label{sec:discretization:mortar_coupling}

Similar to the \btsvc method introduced in \cite{Steinbrecher2020}, we employ a mortar-type coupling approach for all three positional coupling variants, \ie the Lagrange multiplier field~$\lagrangePos$ introduced in \Cref{sec:problem_formulation:btssc} is also interpolated with finite element shape functions, \cf \cite{BenBelgacem1999, Popp2009, Wohlmuth2000, Steinbrecher2020}.
The discrete Lagrange multiplier field is defined along the discretized beam centerline.
Its finite element interpolation reads
\begin{equation}
\lagrangePosh = \sumlagrange{\NlagrangePosni(\xibeam) \qlagrangePosn},
\end{equation}
where $\nlagrange$ is the total number of Lagrange multiplier nodes,~$\NlagrangePosni$ is the discrete Lagrange multiplier shape function of node $\indexlagrange$, and $\qlagrangePosn \in \R{3}$ is the Lagrange multiplier at node~$\indexlagrange$.
Although defined along the beam centerline, there is no requirement for the Lagrange multiplier shape functions to match the shape functions used for interpolation of the beam centerline.
Even the number of nodes can differ, \ie $\nlagrange \ne \nbeam$.

The choice of Lagrange multiplier basis functions is important for the mathematical properties of the resulting discretized system.
Generally speaking, the Lagrange multiplier interpolations must fulfill an $\infsup$ condition to guarantee stability of the mixed finite element method.
We circumvent the $\infsup$ stability condition by employing a penalty regularized Lagrange multiplier approach.
Detailed discussions regarding this topic can be found in \cite{Steinbrecher2020} for the purely positional coupling \btsvc and in \cite{Steinbrecher2022} for rotational coupling \btsvr.
The extensive studies and discussions in these works show that a linear interpolation of the Lagrange multipliers combined with a node-wise weighted penalty regularization generally leads to a stable finite element formulation of the coupled problem, \ie undesirable effects such as contact locking are avoided.
Instabilities might only occur if the beam finite elements become significantly shorter than the surface finite elements.
However, as discussed in \cite{Steinbrecher2020,Steinbrecher2022}, such \btss element size ratios are typically not relevant for the envisioned scope of applications.

\subsubsection{Consistent positional coupling (\nameConsistent)}

Inserting the finite element interpolations into the first term of \eqref{eq:problem_formulation:trans_weak_form} yields the discrete variation of the \nameConsistent coupling constraints,
\begin{equation}
\label{eq:discretization:virtual_work_constraints}
\begin{split}
\dWposLambdah = &
\sumbeam{\sumlagrange{
	\dqlagrangePosn\tr
	\underbrace{
	\intCouplingh{
	\NlagrangePosni
	\Nbeamn
	}}_{\Dn} \qxbeamn
}}
\\ &
-
\sumsurface{\sumlagrange{
	\dqlagrangePosn\tr
	\underbrace{
	\intCouplingh{
	\NlagrangePosni
	\Nsolidn
	} \matI^{3\times3}}_{\Mn} \qxsolidn
}}
\\ &
-
\sumlagrange{
	\dqlagrangePosn\tr
	\underbrace{
	\intCouplingh{
	\NlagrangePosni
	\surfaceNormalDistanceULO
	\surfaceNormalh
	}}_{\qsurfn}
}
\end{split}
\end{equation}
Here, two local matrices with mass matrix-like structure can be identified: $\Dn \in \R{3\times6}$ and $\Mn \in \R{3\times3}$, \ie the so-called mortar matrices.
Furthermore, the abbreviation $\qsurfn \in \R{3\times 1}$ is introduced, referring to the integral of the surface normal distance weighted with the Lagrange multiplier shape function of the Lagrange multiplier node $\indexlagrange$.
Again, inserting the finite element interpolations into the second term of \eqref{eq:problem_formulation:trans_weak_form} yields the  discrete virtual work of the coupling forces,
\begin{equation}
\label{eq:discretization:virtual_work_forces}
\begin{split}
\dWposCh = &
\sumbeam{\sumlagrange{
	\br{\Dn \dqbeamn}\tr \qlagrangePosn
}}
\\ &
-
\sumsurface{\sumlagrange{
	\br{\Mn \dqsolidn}\tr \qlagrangePosn
}}
\\ &
-
\sumsurface{\sumlagrange{
	\br[4]{\,
	\underbrace{
	\intCouplingh{
	\surfaceNormalDistanceULO
	\dsurfaceNormalh
	\NlagrangePosni
	}}_{
		-\Qsurfn \dqsolidn
	}\,}\tr \qlagrangePosn
}}
\end{split}
\end{equation}
where the abbreviation $\Qsurfn = -\pfrac{\qsurfn}{\qsolidn}$ is introduced.
With equations \Cref{eq:discretization:virtual_work_constraints,eq:discretization:virtual_work_forces} the discretized global virtual work of the coupling contributions reads
\begin{equation}
\label{eq:discretization:btss-virtual-work}
\dcouplingPotentialPosMortarh
=
\matrix{\dQsolid\tr & \dQbeam\tr & \dQlagrangePos\tr}
\matrix{
	\br{-\M + \Qsurfg}\tr \QlagrangePos
	\\
	\D\tr \QlagrangePos
	\\
	\D \Qxbeam - \M \Qxsolid - \qsurfg
}.
\end{equation}
Here, $\D \in \R{3\nlagrange \times 6\nbeam}$, $\M \in \R{3\nlagrange \times 3\nsurface}$, $\qsurfg \in \R{3\nlagrange \times 1}$ and~$\Qsurfg \in \R{3\nlagrange \times 3\nsurface}$ are the globally assembled matrices and vector of the previously defined local ones.
The following residual vectors can be identified in \eqref{eq:discretization:btss-virtual-work}
\begin{equation}
\matrix{
	\RcsolidPos
	\\
	\RcbeamPos
	\\
	\RcPos
}_\text{\nameConsistentShort}
=
\matrix{
	\br{-\M + \Qsurfg}\tr \QlagrangePos
	\\
	\D\tr \QlagrangePos
	\\
	\D \Qxbeam - \M \Qxsolid - \qsurfg
}.
\end{equation}
Here, the abbreviations $\RcsolidPos$ and $\RcbeamPos$ are the coupling residual force vectors acting on the structure and beam degrees of freedom, respectively, and $\RcPos$ is the residual vector of the constraint equations.
The residual vectors for the positional coupling conditions are added to those of the uncoupled system \eqref{eq:discretization:system_uncoupled}.
A linearization of the coupling residuum vectors with respect to the discrete degrees of freedom is required for the \nr algorithm used to solve the nonlinear system of equations resulting from the discretization process.
The linearization of the positional coupling contributions reads:
\begin{equation}
\label{eq:discretization:bts-full-global-residuum}
\begin{split}
\lin\br{
\matrix{\RcsolidPos \\ \RcbeamPos \\ \RcPos}_\text{\nameConsistentShort}
}
= &
\matrix{\vectO \\ \vectO \\ \RcPos}_\text{\nameConsistentShort}
\\ & \hspace{-10mm}
+
\matrix{
\Qsurfsurfg & \matO & -\M\tr + \Qsurfg\tr \\
\matO & \matO & \D\tr \\
-\M + \Qsurfg & \D & \matO
}
\matrix{\DQsolid \\ \DQbeam \\ \QlagrangePos},
\end{split}
\end{equation}
where the abbreviation $\Qsurfsurfg = \pfrac{\br{\Qsurfg\tr \QlagrangePos}}{\Qsolid}$ is introduced.

In practice, all integrals are numerically evaluated using segment-based integration along the beam centerline, which avoids integration over discontinuities, \cf \cite{Farah2015, Steinbrecher2020}.
Each subsegment is integrated using \gale quadrature with a fixed number of integration points for all coupling terms, which is required to ensure conservation of linear and angular momentum, \cf \Cref{sec:discretization:conservation_laws}.
Segment-based integration yields an accurate numerical evaluation of the coupling integrals and allows for the resulting coupling scheme to pass patch test-like problems, \cf \cite{Steinbrecher2020}.
Furthermore, all derivatives explicitly stated in the discrete equations are evaluated using forward automatic differentiation (\fad), \cf \cite{Korelc2016}, using the Sacado software package \cite{SacadoWebsite}, which is part of the Trilinos project \cite{TrilinosWebsite}.

\subsubsection{Forced reference configuration coupling (\nameReference)}
\label{sec:discretization:forced_ref}

By neglecting the normal distance $\surfaceNormalDistanceULO$, the \nameReference variant of the positional coupling conditions \eqref{eq:problem_formulation:position_coupling_ref} simplifies the coupling equations \eqref{eq:problem_formulation:position_coupling}, such that the surface normal vector does not appear in the coupling equations anymore.
The discrete coupling terms for the \nameReference variant read
\begin{equation}
\label{eq:discretization:btssc_forced_reference_res}
\matrix{\RcsolidPos \\ \RcbeamPos \\ \RcPos}_{\text{\nameReferenceShort}}
=
\matrix{
	-\M\tr \QlagrangePos
	\\
	\D\tr \QlagrangePos
	\\
	\D \Qxbeam - \M \Qxsolid
}.
\end{equation}
It becomes clear, that the constraint equations in the reference configuration are only fulfilled if $\D \QXbeam - \M \QXsolid = \matO$.
Otherwise, this coupling variant leads to initial (coupling) stresses in the system.
The influence of the initial stresses within the \nameReference variant is analyzed in \Cref{sec:examples}.
The linearization of the coupling terms \eqref{eq:discretization:btssc_forced_reference_res} reads
\begin{equation}
\label{eq:discretization:btssc_forced_reference}
\begin{split}
\lin\br{
\matrix{\RcsolidPos \\ \RcbeamPos \\ \RcPos}_{\text{\nameReferenceShort}}
}
= &
\matrix{\vectO \\ \vectO \\ \RcPos}_{\text{\nameReferenceShort}}
\\ &
+
\matrix{
\matO & \matO & -\M\tr \\
\matO & \matO & \D\tr \\
-\M & \D & \matO
}
\matrix{\DQsolid \\ \DQbeam \\ \QlagrangePos}.
\end{split}
\end{equation}

\subsubsection{Displacement coupling (\nameDisplacement)}
Another alternative positional coupling variant is the \nameDisplacement variant \eqref{eq:problem_formulation:position_coupling_disp}.
Therein, the normal distance between the beam and the surface is neglected and the displacements are directly coupled to each other.
The discrete coupling terms for the \nameDisplacement variant read:
\begin{equation}
\label{eq:discretization:btssc_displacement_res}
\matrix{\RcsolidPos \\ \RcbeamPos \\ \RcPos}_{\text{\nameDisplacementShort}}
=
\matrix{
	-\M\tr \QlagrangePos
	\\
	\D\tr \QlagrangePos
	\\
	\D \Qbeam - \M \Qsolid
}.
\end{equation}
In this case, the coupling constraints are fulfilled in the reference configuration and there are no initial stresses in the system.
However, this variant violates conservation of angular momentum.
Again, the influence of this violation within the \nameDisplacement variant is analyzed in \Cref{sec:examples}.
The linearization of the coupling terms \eqref{eq:discretization:btssc_displacement_res} reads
\begin{equation}
\label{eq:discretization:btssc_displacement}
\begin{split}
\lin\br{
\matrix{\RcsolidPos \\ \RcbeamPos \\ \RcPos}_{\text{\nameDisplacementShort}}
}
= &
\matrix{\vectO \\ \vectO \\ \RcPos}_{\text{\nameDisplacementShort}}
\\ &
+
\matrix{
\matO & \matO & -\M\tr \\
\matO & \matO & \D\tr \\
-\M & \D & \matO
}
\matrix{\DQsolid \\ \DQbeam \\ \QlagrangePos}.
\end{split}
\end{equation}

\begin{remark}
\label{rem:discretization:mesh_initialization}
A very similar problem occurs for \sutsu mesh tying problems in the case of general curved interfaces.
To guarantee the conservation of angular momentum, it is common practice to perform a mesh initialization procedure, \cf \cite{Puso2004}.
The mesh initialization slightly relocates the reference position of the slave nodes (in the case of \btss problems, the beam nodes)~$\QXbeam$, such that the (non-linear) condition~$\D \QXbeam - \M \QXsolid = \matO$ is fulfilled (in \cite{Puso2004}, the non-linear mesh initialization is not solved exactly, but approximated with a single linear step).
For the presented \btsstrans method, such a mesh initialization would mean that both presented simplifications \nameReference and \nameDisplacement are identical.
However, in the \sutsu case the space continuous interfaces are usually matching, thus the mesh initialization of the discretized system only marginally affects the overall solution.
This is not the case for \btss coupling problems.
For example, in many situations it is sensible for the beam centerline to be offset of the coupling surface in surface normal direction.
In such cases, the mesh initialization procedure of the beam reference configuration will lead to a drastically different system behavior.
\end{remark}

\begin{remark}
If the discretized beam centerline lies exactly on the discretized surface, the three presented variants of the global system equations \Cref{eq:discretization:bts-full-global-residuum,eq:discretization:btssc_forced_reference,eq:discretization:btssc_displacement} are identical, \ie $\Qsurfg = \matO$ and $\D \QXbeam - \M \QXsolid = \matO$.
However, with the employed Lagrange polynomial interpolation for the solid finite elements and the third-order Hermitian interpolation for the beam finite elements, a matching mesh for beam and surface discretizations is only possible in case of planar surfaces.
\end{remark}

\subsection{Mortar-type coupling of rotations (\btssrot)}

The rotational coupling between beam \cs and surface (\btssrot) is based on the \btsvr coupling method presented in \cite{Steinbrecher2022}.
Therein, a mortar-type approach is employed to weakly enforce the rotational coupling constraints \eqref{eq:problem_formulation:rotation_coupling}.
The linearization of the global residuum vectors for rotational coupling reads,
\begin{equation}
\label{eq:discretization:system_rotational_coupling}
\lin\br{
\matrix{\RcsolidRot \\ \RcbeamRot \\ \RcRot}
}
=
\matrix{\vectO \\ \vectO \\ \RcRot}
+
\matrix{
	\Qcss & \Qcsb & \Qcsl \\
	\Qcbs & \Qcbb & \Qcbl \\
	\Qcls & \Qclb & \matO
}
\matrix{\DQsolid \\ \DQbeamrot \\ \QlagrangeRot}.
\end{equation}
Here, $\QcPlace$ are the rotational coupling matrices and $\QlagrangeRot$ are the Lagrange multipliers enforcing the rotational coupling constraints.
Furthermore, $\RcsolidRot$ and $\RcbeamRot$ are the rotational coupling residual force vectors associated with the structure and beam degrees of freedom, respectively, and $\RcRot$ is the residual vector of the rotational constraint equations.
For the derivation of the rotational coupling terms the interested reader is referred to our previous publication \cite[Section 6.3]{Steinbrecher2022}

\subsection{Combined mortar-type coupling and penalty regularization (\btssfull)}

\subsubsection{Coupling equations}
\label{sec:discretization:combined_system}

In this section the global system for the \btssfull problem is assembled and subsequently regularized.
The \btssfull problem consists of the following individual parts: the uncoupled beam and structure problem, the positional coupling (\btsstrans) and the rotational coupling (\btssrot).
In \Cref{sec:discretization:mortar_coupling} three different variants of \btsstrans are presented: \nameConsistent, \nameReference and \nameDisplacement.
Depending on the employed variant, the corresponding \btssfull problem is referred to as \btssfullcons, \btssfullref and \btssfulldisp.
\btssfullcons is the most general of the presented variants, \ie the equations for the other variants are more or less simplifications thereof.
In \Cref{sec:examples}, \btssfullcons will be identified as the superior variant with respect to the accuracy of the results.
Therefore, and for the sake of brevity, the following derivations are only presented for the fully coupled and consistent \btssfullcons variant.
To improve readability, the subscript $\placeholder_\text{\nameConsistentShort}$ will be omitted going further.

Combining the individual contributions to the \btssfullcons problem, \ie the uncoupled system \eqref{eq:discretization:system_uncoupled}, the positional \nameConsistent coupling terms \eqref{eq:discretization:bts-full-global-residuum} and the rotational \btssrot coupling terms \eqref{eq:discretization:system_rotational_coupling}, yields the following global system of equations:
\begin{equation}
\label{eq:discretization:system_btssc}
\begin{split}
\matrix{
	\Kss + \Qsurfsurfg + \Qcss & \matO & \Qcsb & -\M\tr + \Qsurfg\tr & \Qcsl \\
	\matO & \Krr & \matO & \D\tr & \matO \\
	\Qcbs & \matO & \Ktt + \Qcbb & \matO & \Qcbl \\
	-\M + \Qsurfg & \D & \matO & \matO & \matO \\
	\Qcls & \matO & \Qclb & \matO & \matO
}
\\ \cdot
\vector{\DQsolid \\ \DQbeam \\ \DQbeamrot \\ \QlagrangePos \\ \QlagrangeRot}
= \begin{bmatrix}
-\Rsolid \\
-\RbeamPos \\
-\RbeamRot \\
-\RcPos \\
-\RcRot
\end{bmatrix}.
\end{split}
\end{equation}

\subsubsection{Penalty regularization}

Enforcing the coupling conditions with Lagrange multipliers results in a mixed formulation, \ie the Lagrange multipliers are additional global unknowns.
This leads to a saddle point-type structure of the global system \eqref{eq:discretization:system_btssc}.
A direct solution of this global system introduces certain drawbacks, \eg an increased system size and possible linear solver issues due to the saddle point-type structure.
A weighted penalty regularization has proven to be an efficient and reasonably accurate approach to circumvent the aforementioned drawbacks for 1D-3D \btv coupling problems, \cf \cite{Steinbrecher2020,Steinbrecher2022}.
Therefore, the resulting global system \eqref{eq:discretization:system_btssc} will also be approximated with a penalty regularization.
For the rotational coupling constraints the same penalty relaxation as in \cite{Steinbrecher2022} will be employed, \ie~$\QlagrangeRot = \penRot \ScalingMatrixRot\inv \RcRot$.
Therein, $\penRot \in \R{+}$ is a scalar penalty parameter and $\ScalingMatrixRot$ is a diagonal scaling matrix to account for the non-uniform weighting of the constraint equations.
A similar relaxation is employed for the positional coupling constraints,
\begin{equation}
\QlagrangePos = \penPos \ScalingMatrix\inv \RcPos.
\end{equation}
Again, $\penPos \in \R{+}$ is a scalar penalty parameter and $\ScalingMatrix$ is a global diagonal scaling matrix.
The global scaling matrix is assembled from the nodal scaling matrices $\scalingMatrix^{(i,i)}$ for Lagrange multiplier node $i$, \cf \cite{Steinbrecher2020,Steinbrecher2022}, \ie
\begin{equation}
\label{eq:kappa}
\scalingMatrix^{(i,i)} = \intCouplingh{\NlagrangePosni} \matI^{3\times 3}.
\end{equation}

The penalty regularization introduces two additional system parameters, $\penPos$ and $\penRot$.
This leaves the important question on how to choose these two parameters.
Obviously, choosing the penalty parameters too high can lead to an ill-conditioned system matrix and subsequent issues with the numerical solution procedure, as well as to contact locking effects \cite{Steinbrecher2020}.
Moreover, also from a mechanical point of view, an infinitely large penalty parameter is not desirable.
This is because in the real physical problem the beam \cs comes with a certain deformability.
However, the employed beam theory introduces the assumption of rigid \cs{s}.
Therefore, the penalty parameter is no longer a pure mathematical tool of constraint enforcement, but it also has a physical meaning, \ie it represents the beam \cs stiffness.
Similar observations can be made in the case of \btb contact, \cf \cite{Meier2016a}.
Going further, one could define the penalty parameter based on continuum mechanical analysis of the \cs deformability and stiffness.
However, since our primary interest is the regularization of \eqref{eq:discretization:system_btssc}, the following rule of thumb for choosing the two penalty parameters can be given: the positional penalty parameter should be in the range of the Young's modulus of the beam, \ie $\penPos \approx E_{\letterbeam}$, and the rotational parameter should be in the range of the Young's modulus of the beam scaled with the square of the \cs radius, \ie $\penRot \approx E_{\letterbeam} R^2$.
In practice, this does not lead to an unphysically large violation of the coupling constraints, and contact locking has not been observed in combination with a linear interpolation of the Lagrange multiplier field, \cf \cite{Steinbrecher2020}.

The relaxation of the penalty constraints defines the Lagrange multipliers as functions of the displacements, \ie they are no longer independent degrees of freedom of the system and can be removed from the global system of equations \eqref{eq:discretization:system_btssc}:
\begin{equation}
\matrix{
\mat{A}_{ss} & \mat{A}_{s\letterbeamcenterline} & \mat{A}_{s\letterbeamrot} \\
\mat{A}_{\letterbeamcenterline s} & \mat{A}_{\letterbeamcenterline\letterbeamcenterline} & \mat{A}_{\letterbeamcenterline\letterbeamrot} \\
\mat{A}_{\letterbeamrot s} & \mat{A}_{\letterbeamrot\letterbeamcenterline} & \mat{A}_{\letterbeamrot\letterbeamrot}
}
\vector{\DQsolid \\ \DQbeam \\ \DQbeamrot}
=
\begin{bmatrix}
\mat{B}_{s}
\\
\mat{B}_{\letterbeamcenterline}
\\
\mat{B}_{\letterbeamrot}
\end{bmatrix}.
\end{equation}
Therein, the following abbreviations have been introduced for improved readability:
\begin{equation}
\begin{split}
\mat{A}_{ss} &= \Kss + \Qsurfsurfg + \Qcss \\ & \quad + \penPos \br{-\M+\Qsurfg}\tr \ScalingMatrix\inv \br{-\M+\Qsurfg} \\ & \quad + \penRot \Qcsl \ScalingMatrixRot\inv \Qcls\\
\mat{A}_{s\letterbeamcenterline} &= \penPos \br{-\M+\Qsurfg}\tr \ScalingMatrix\inv \D \\
\mat{A}_{s\letterbeamrot} &= \Qcsb + \penRot \Qcsl \ScalingMatrixRot\inv \Qclb\\
\mat{A}_{\letterbeamcenterline s} &= \penPos \D\tr \ScalingMatrix\inv \br{-\M+\Qsurfg}\\
\mat{A}_{\letterbeamcenterline\letterbeamcenterline} &= \Krr + \penPos \D\tr \ScalingMatrix\inv \D\\
\mat{A}_{\letterbeamcenterline\letterbeamrot} &= \Krt\\
\mat{A}_{\letterbeamrot s} &= \Qcbs + \penRot \Qcbl \ScalingMatrixRot\inv \Qcls\\
\mat{A}_{\letterbeamrot\letterbeamcenterline} &= \Ktr\\
\mat{A}_{\letterbeamrot\letterbeamrot} &= \Ktt + \Qcbb + \penRot \Qcbl \ScalingMatrixRot\inv \Qclb \\
\mat{B}_{s} &= -\Rsolid - \penPos \br{-\M + \Qsurfg}\tr \ScalingMatrix\inv \RcPos \\ & \quad - \penRot \Qcsl \ScalingMatrixRot\inv  \RcRot \\
\mat{B}_{\letterbeamcenterline} &= -\Rbeam - \penPos \D\tr \ScalingMatrix\inv \RcPos \\
\mat{B}_{\letterbeamrot} &= -\RbeamRot - \penRot \Qcbl \ScalingMatrixRot\inv  \RcRot
.\end{split}
\end{equation}

\subsubsection{Conservation properties}
\label{sec:discretization:conservation_laws}

In this section, the proposed \btssfullcons scheme shall be analyzed with respect to conservation of linear momentum and angular momentum.
For the rotational coupling constraints conservation of angular momentum (there is no linear momentum introduced by the rotational coupling constraints) is shown in \cite{Steinbrecher2022}.
Therefore, it is sufficient to analyze the \nameConsistent scheme in this section.

In the context of \sutsu problems this has been discussed in detail, \eg \cite{Puso2004,Puso2004a,Puso2004b,Popp2010}.
For \sutsu coupling (mesh tying) problems it has been shown that conservation of linear momentum and angular momentum is satisfied by the semi-discrete mesh tying formulation, \cf \cite{Puso2004}.
However, the proposed mixed-dimensional \nameConsistent scheme differs in two important aspects from classical \sutsu coupling problems: (i) The coupling constraints are formulated with the current positions instead of the displacements, and, more importantly, contain the surface normal vector.
(ii) The nodal degrees of freedom for the beam nodes contain the positions as well as the centerline tangents.
Therefore, a discussion on conservation of linear momentum and angular momentum of the \nameConsistent variant is warranted.
In the following considerations the \nameConsistent is analyzed, as the implications for \nameReference can be directly obtained by applying the respective simplifications.
In the case of \nameDisplacement, it is shown in \Cref{sec:variants_disp} that already the space continuous coupling terms do not conserve angular momentum.

As discussed in \cite{Puso2004}, conservation of linear momentum can be guaranteed if the discretized virtual work of the coupling forces vanishes for a constant virtual displacement~$\virtDisp \ne \tnsO$.
In that case, the nodal displacement weighting functions become~$\dqsolidn=\virtDisp,\, \indexsolid=1,...,\nsurface$ and~$\dqbeamer=\virtDisp,\, \indexbeam=1,...,\nbeam$.
Since the virtual displacement is constant, the variation of the beam centerline tangents vanishes, \ie~$\dqbeamet = \tnsO, \, \indexbeam=1,...,\nbeam$.
Insertion into \eqref{eq:discretization:virtual_work_forces} yields
\begin{equation}
\label{eq:discretization:linear_momentum_conservation}
\begin{split}
&
\sumlagrange{ \brackets{(}{.}{
\sumbeam{
	\br{\Dn \matrix{\virtDisp \\ \tnsO}}\tr
}
- \sumsurface{ \br{
	\Mn \virtDisp}\tr}
}}
\\ & \qquad
\brackets{.}{)}{
- \intCouplingh{
		\surfaceNormalDistanceULO
		\dsurfaceNormalh
		\NlagrangePosni}
}
\qlagrangePosn = 0.
\end{split}
\end{equation}
The variation of the surface normal vector vanishes for a constant virtual displacement field, \ie $\dsurfaceNormalh = \tnsO$.
Furthermore, since $\virtDisp$ is non-zero, the condition \eqref{eq:discretization:linear_momentum_conservation} is only satisfied if
\begin{equation}
\br{
\sumbeam{
	\intCouplingh{
	\NlagrangePosni
	\Nbeamr
	}}
-
\sumsurface{
	\intCouplingh{
	\NlagrangePosni
	\Nsolidn
	}
}} \qlagrangePosn = \tnsO.
\end{equation}
With the partition of unity property of $\Nbeamr$ and $\Nsolidn$, \ie $\sumbeam{\Nbeamr} = 1$ and $\sumsurface{\Nsolidn} = 1$, the condition for conservation of linear momentum further simplifies to
\begin{equation}
\sumlagrange{ \br{
	\intCouplingh{
	\NlagrangePosni
	}
	-\intCouplingh{
	\NlagrangePosni
}}} = 0.
\end{equation}
Obviously this property is fulfilled if the integrals are evaluated exactly.
In the case of numerical integration the property is fulfilled if the same numerical integration procedure is used for both integrals.
At this point it is important to point out that the two integrals originally arise from the evaluation of $\Dn$ and $\Mn$.
As mentioned in \Cref{sec:discretization:mortar_coupling} a segment-based integration with a fixed number of Gauss-points is performed, therefore, the discrete \nameConsistent scheme exactly conserves linear momentum.

In a similar fashion, conservation of angular momentum can be guaranteed, if the virtual work of the coupling forces vanishes for a constant virtual rotation $\virtRot \ne \tnsO$ (for simplicity, the origin is assumed to be the center of the virtual rotation).
With that assumption, the nodal virtual displacements of solid and beam are $\dqsolidn=\virtRot \times \qxsolidn,\, \indexsolid=1,...,\nsurface$ and~$\dqbeamer=\virtRot \times \qxbeamer,\, \indexbeam=1,...,\nbeam$.
The variation of the nodal beam tangent vectors reads~$\dqbeamet=\virtRot \times \qxbeamet,\, \indexbeam=1,...,\nbeam$, \cf \cite{Meier2019}.
Insertion into \eqref{eq:discretization:virtual_work_forces} yields
\begin{equation}
\label{eq:discretization:angular_momentum_conservation_1}
\begin{split}
&
\sumlagrange{ \brackets{[}{.}{
\sumbeam{\brackets{(}{.}{
	 \br{\virtRot \times\qxbeamer}\tr
	 \intCouplingh{
	 	\NlagrangePosni
	 	\Nbeamr
	 	}}}}}
\\ & \qquad
+ \brackets{.}{)}{\br{\virtRot \times\qxbeamet}\tr
	 \intCouplingh{
	 	\NlagrangePosni
	 	\Nbeamt
	 	}}
\\ & \qquad
-
\sumsurface{
	\br{\virtRot \times\qxsolidn}\tr \intCouplingh{\NlagrangePosni \Nsolidn}
	}
\\ & \qquad
\brackets{.}{]}{
	-\intCouplingh{
		\surfaceNormalDistanceULO
		\dsurfaceNormalh\tr
		\NlagrangePosni
		}
} \qlagrangePosn = 0.
\end{split}
\end{equation}
In the case of a constant virtual rotation, the variation of the normal vector can be expressed as $\dsurfaceNormalh = \virtRot \times \surfaceNormalh$.
Since~$\virtRot$ is non-zero, the condition \eqref{eq:discretization:angular_momentum_conservation_1} is only fulfilled if
\begin{equation}
\label{eq:discretization:angular_momentum_conservation}
\begin{split}
&\sumlagrange{ \brackets[4]{[}{.}{
\sumbeam{\underbrace{\br{
	\intCouplingh{
		\NlagrangePosni
	 	\Nbeamr
	 	}
	 \qxbeamer
	 +
	 \intCouplingh{
	 	\NlagrangePosni
	 	\Nbeamt
	 	}
	 \qxbeamet
	 }}_{\Dn \qxbeamn}
}}}
\\& \qquad
-
\sumsurface{
	\underbrace{\intCouplingh{\NlagrangePosni \Nsolidn} \matI^{3\times 3}}_{\Mn\tr} \qxsolidn
	}
\\& \qquad
\brackets[4]{.}{]}{
	-\underbrace{\intCouplingh{
		\surfaceNormalDistanceULO
		\surfaceNormalh
		\NlagrangePosni
		}}_{\qsurfn}
} \times \qlagrangePosn = 0.
\end{split}
\end{equation}
This condition can be reformulated and written in global form
\begin{equation}
\D \qxbeam - \M \qxsolid - \qsurfg = \matO.
\end{equation}
These are simply the coupling constraints for \nameConsistent, \ie if the coupling constraints, \cf last row in \eqref{eq:discretization:btss-virtual-work}, are fulfilled, the coupling scheme preserves angular momentum.
In the present work, the coupling constraints are enforced with a node-wise weighted penalty regularization, which results in a slight violation of the coupling constraints.
However, the resulting regularized problem still preserves angular momentum.
To demonstrate this, we state the penalty regularization for a Lagrange multiplier at node $\indexlagrange$:
\begin{equation}
\label{eq:discretization:angular_momentum_conservation_penalty}
\begin{split}
\qlagrangePosn = & \, \penPos \br{\intCouplingh{\NlagrangePosni} \matI^{3\times 3}}\inv
\\
&
\br{
	\sumbeam{\Dn \qxbeamn} - \sumsurface{\Mn \qxsolidn} - \qsurfn
}.
\end{split}
\end{equation}
When inserting \eqref{eq:discretization:angular_momentum_conservation_penalty} into \eqref{eq:discretization:angular_momentum_conservation} it is obvious that the condition for conservation of angular momentum is also fulfilled for the regularized problem, as the cross product of two parallel vectors vanishes.

\section{Examples}
\label{sec:examples}

In this section, we present several numerical examples to evaluate the different \btss variants proposed in this work.
In \Cref{sec:examples:patch}, we consider 2D surfaces of 3D solid bodies as coupling surfaces (\ie \btsso scenario).
In \Cref{sec:examples:coupling,sec:examples:artery}, the coupling surface is part of a so-called solid shell discretization.
There, the underlying structural elements are eight-noded hexahedral elements, \cf \cite{Vu-Quoc2003a,Vu-Quoc2003b}.
From a geometrical point of view, the coupling surface can therefore be classified as a 2D surface of a 3D solid (\ie \btsso scenario).
In \Cref{sec:examples:shell}, the surface is modeled using a reduced-dimensional 2D \kl shell formulation (\ie \btssh scenario).
All numerical examples are set up using the open source beam finite element pre-processor MeshPy~\cite{MeshPyWebsite} and are simulated with the open source multi-physics research code 4C~\cite{FourCWebsite}.

\subsection{Constant stress transfer}
\label{sec:examples:patch}

\begin{figure}
\centering
\includegraphics[scale=1]{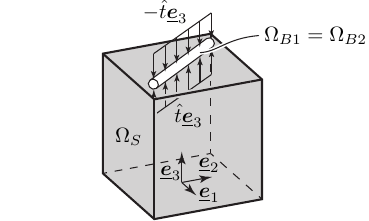}
\caption{%
Constant stress transfer -- problem setup.
The two beams $\domainBeam[1]$ and $\domainBeam[2]$ occupy the same spatial position.
}
\label{fig:examples:patch_test:problem}
\end{figure}

In this first example the ability of the \btssfull coupling method to transfer a constant stress state is investigated.
This example is inspired by classical patch tests for solid mechanics, \cf \cite{Taylor1986}.
Similar examples are presented in \cite{Steinbrecher2020} for the \btsvc method and in \cite{Steinbrecher2022} for the \btsvrc method.
\Cref{fig:examples:patch_test:problem} illustrates the problem setup, which consists of a solid block~$\Omega^{S}$ ($\Esolid = \unit[1]{N/m^2}$, $\nusolid = 0$) with the dimensions $\unit[1]{m} \times \unit[1]{m} \times \unit[1.2]{m}$.
The center of the bottom face is located at the origin of the coordinate system.
No external loads are applied to the solid and the bottom face is fixed in all spatial directions.
At the top face the solid surface is coupled to two beams $\domainBeam[1]$ and $\domainBeam[2]$ ($\radius = \unit[0.05]{m}$, $\Ebeam=\unit[100]{N/m^2}$, $\nubeam = 0$).
The two beams share the same spatial position and are loaded with opposing line loads in $\ez$ direction.
The magnitude of the line loads is $\hat{t}=\unit[0.025]{N/m}$.
Note that this verification example is designed in a manner such that the two beams do not interact directly, \eg via mechanical contact interaction, all loads are transferred through the solid domain via the \btssfull coupling method.
For the space continuous problem, the coupling forces resulting from two identical beams loaded with opposing line loads exactly balance each other, \ie the net coupling force transferred to the surface vanishes and, thus, the analytical solution for the displacement field of the beams and the solid is $\tns{u} = \tnsO$.
This example shall verify the ability of the three \btssfull variants proposed in \Cref{sec:variants}, \ie \btssfullcons, \btssfullref and \btssfulldisp, to exactly represent this analytical solution using an arbitrarily coarse discretization, \ie the ability of the coupling method to transfer a constant stress state across non-matching mixed-dimensional interface meshes.

\begin{figure*}
\centering
\subfigure[\btssfullcons]{\includegraphicsdpi{300}{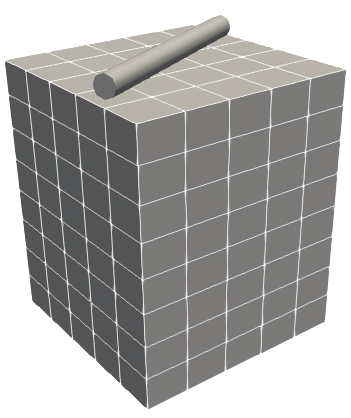}}
\hfil
\subfigure[\btssfullref]{\includegraphicsdpi{300}{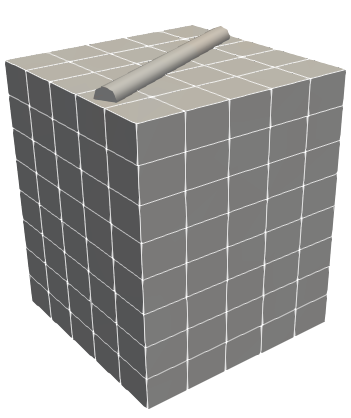}}
\hfil
\subfigure[\btssfulldisp]{\includegraphicsdpi{300}{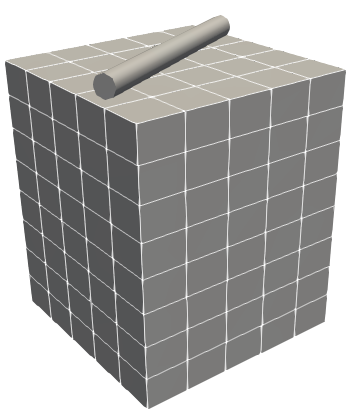}}
\hfil
{\normalsize\input{figures/examples_patch_test_plane_color_bar.tex}}
\caption{%
Constant stress transfer -- results for straight beams and various coupling variants.
The solids are discretized with \hex{8} solid finite elements.
The second Piola-Kirchhoff stress $S_{33}$ is shown in the solid and the curvature $\kappa$ at the middle of each beam element.
Note that the two beams $\domainBeam[1]$ and $\domainBeam[2]$ occupy the same spatial domain in the undeformed reference configuration.
The gray color in the contour plot indicates a zero value up to machine precision.
}
\label{fig:examples:patch_test:results_plane_variants}
\end{figure*}

\begin{figure*}
\centering
\subfigure[\hex{8}, \hex{20}, \hex{27}]{\includegraphicsdpi{300}{figures/examples_patch_test_plane_hex8.png}}
\hfil
\subfigure[\tet{4}, \tet{10}]{\includegraphicsdpi{300}{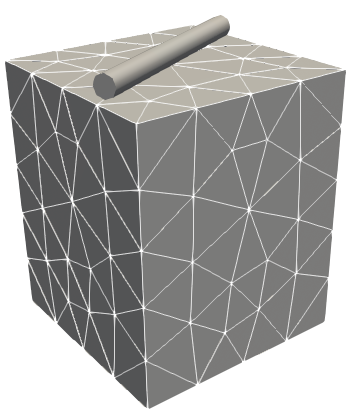}}
\hfil
{\normalsize\input{figures/examples_patch_test_plane_color_bar.tex}}
\caption{%
Constant stress transfer -- results for straight beams and various solid finite element discretizations.
The coupling is modeled with the \btssfullcons coupling variant.
The second Piola-Kirchhoff stress~$S_{33}$ is shown in the solid and the curvature~$\kappa$ at the middle of each beam element.
Note that the two beams~$\domainBeam[1]$ and~$\domainBeam[2]$ occupy the same spatial domain in the undeformed reference configuration.
The gray color in the contour plot indicates a zero value up to machine precision.
}
\label{fig:examples:patch_test:results_plane}
\end{figure*}

The solid block is discretized with first- and second-order hexahedral finite elements (\hex{8}, \hex{20} and \hex{27}) as well as first- and second order tetrahedral finite elements (\tet{4} and \tet{10}).
The beams $B1$ and $B2$ are discretized with 5 and 7 \sr beam finite elements, respectively.
This results in a non-matching mixed-dimensional interface discretization between the beams and the solid.
The Lagrange multipliers for positional and rotational coupling are discretized using first-order Lagrange polynomials and regularized using penalty parameters of~$\penPos = \unit[100]{N/m^2}$ and $\penRot = \unit[0.1]{Nm/m}$.
The results for various coupling variants and \hex{8} elements are illustrated in \Cref{fig:examples:patch_test:results_plane_variants}.
It can be seen that for all considered variants, the second Piola-Kirchhoff stress~$S_{33}$ in the solid and the curvature $\kappa$ in the beam elements are zero up to machine precision, thus exactly representing the analytical solution.
However, the displacement of the two beams in the \btssfullref variant does not vanish, as the beam centerline is forced to lie on the surface, \ie in this example the beams exhibit an offset in negative~$\ez$-direction by a distance of $\radius$.
The results of the constant stress transfer test for the various solid element types are visualized in \Cref{fig:examples:patch_test:results_plane}.
There, the coupling is realized with the \btssfullcons variant.
It can be seen that for all considered solid element types, the stress in the solid and the curvature in the beam match the analytical solution up to machine precision.
This illustrates that the \btssfullcons coupling variant is able to exactly represent a constant stress state between a straight beam and a planar surface for general non-matching discretizations.
The results obtained with \btssfulldisp exactly match the results obtained with \btssfullcons.
In case of the \btssfullref variant, the beams displacement is not zero, but the constant stress state can still be transferred exactly.

\begin{figure*}
\centering
\subfigure[\btssfullcons]{\label{sub1}\includegraphicsdpi{300}{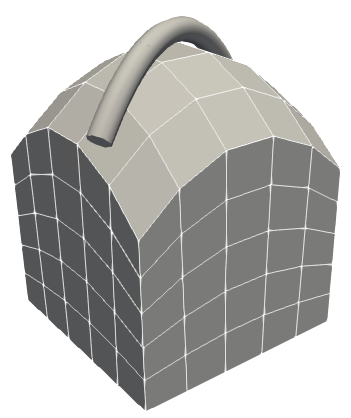}}
\hfil
\subfigure[\btssfullref]{\includegraphicsdpi{300}{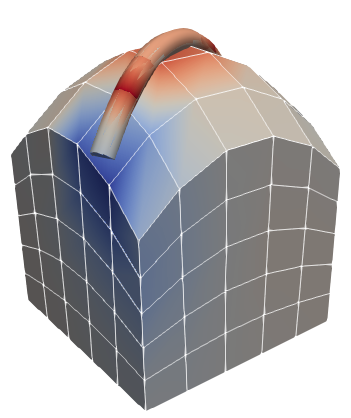}}
\hfil
\subfigure[\btssfulldisp]{\label{sub2}\includegraphicsdpi{300}{figures/examples_patch_test_curved_hex8_displacement.png}}
\hfil
{\normalsize\input{figures/examples_patch_test_curved_displacement_color_bar.tex}}
\caption{%
Constant stress transfer -- results for curved beams and various coupling variants.
The solids are discretized with \hex{8} solid finite elements.
The second Piola-Kirchhoff stress $S_{33}$ is shown in the solid and the curvature $\kappa$ at the middle of each beam element.
Note that the two beams $\domainBeam[1]$ and $\domainBeam[2]$ occupy the same spatial domain in the undeformed reference configuration.
}
\label{fig:examples:patch_test:results_curved_variants}
\end{figure*}

\begin{figure*}
\centering
\subfigure[\hex{8}]{\includegraphicsdpi{300}{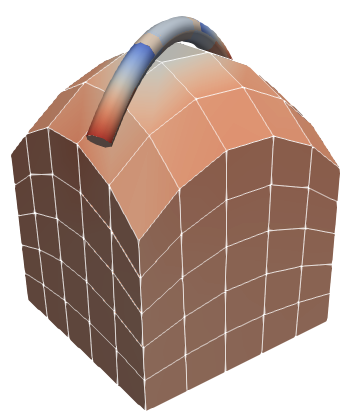}}
\hfil
\subfigure[\hex{20}]{\includegraphicsdpi{300}{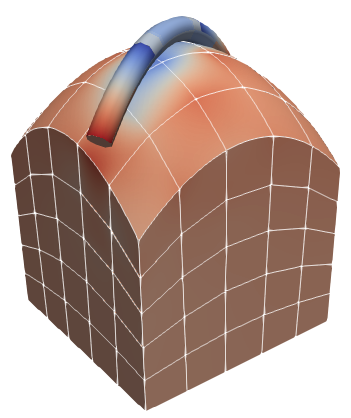}}
\hfil
\subfigure[\hex{27}]{\includegraphicsdpi{300}{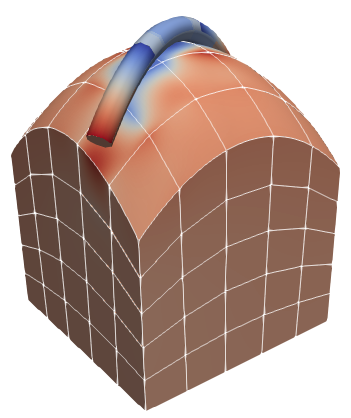}}
\hfil
\subfigure[\tet{4}]{\includegraphicsdpi{300}{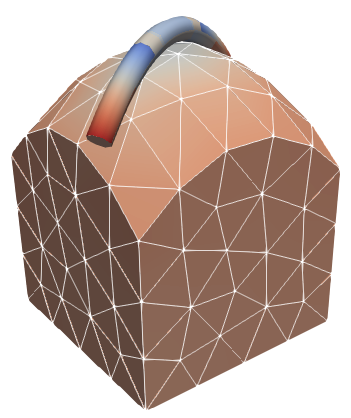}}
\hfil
\subfigure[\tet{10}]{\includegraphicsdpi{300}{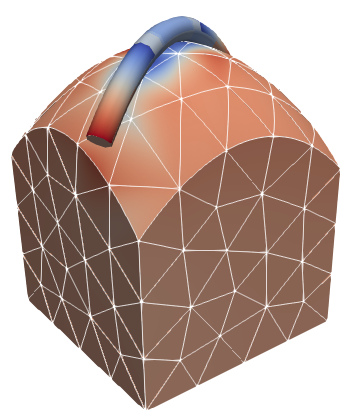}}
\hfil
{\normalsize\input{figures/examples_patch_test_curved_color_bar.tex}}
\caption{%
Constant stress transfer -- results for curved beams and various solid finite element discretizations.
The coupling is modeled with the \btssfullcons coupling variant.
The second Piola-Kirchhoff stress~$S_{33}$ is shown in the solid and the curvature~$\kappa$ at the middle of each beam element.
Note that the two beams~$\domainBeam[1]$ and~$\domainBeam[2]$ occupy the same spatial domain in the undeformed reference configuration.
}
\label{fig:examples:patch_test:results_curved}
\end{figure*}

To make the constant stress transfer test more demanding, the previously presented example is now modified to account for a curved surface contour of the solid described by the position field~$\Xsolid = i \ex + j \ey + f(i,j) \ez$ for $i,j \in [-0.5, 0.5]$, with $f(i, j) = \frac{5}{4}-i^2-j^2$.
The centerlines of the two beams are offset by the beam radius in surface normal direction.
Otherwise, all parameters are the same as in the previous example.
Because of the specific choice of surface curvature, the employed beam centerline interpolation with third-order Hermitian polynomials is not able to exactly represent the space continuous reference geometry of the beam centerline.
This results in a discretization error of the beam centerline interpolation and slightly different arc lengths of the two beams.
In order for the resultants of the two line loads to still be in equilibrium with each other, the load $t$ on beam~$B2$ is scaled with a factor of $0.9995346$ to account for the different beam lengths.
\Cref{fig:examples:patch_test:results_curved_variants} illustrates the results of the constant stress transfer test for the curved surface and the various coupling variants.
It can clearly be seen that the results for \btssfullref do not match the analytical solution.
This is because the beam is forced to lie on the surface.
In case of the planar coupling surface this could be achieved by a rigid body translation of the beams onto the surface.
However, in case of the curved surface, a rigid body translation of the beams can not fulfill the positional coupling equations for \btssfullref.
This also requires a deformation of the beams and the solid, and thus results in a failing constant stress transfer test.
\Cref{fig:examples:patch_test:results_curved} illustrates the results for the \btssfullcons variant in combination with various solid finite element types.
Note the different scaling of the contour plots in \Cref{fig:examples:patch_test:results_curved_variants} compared to \Cref{fig:examples:patch_test:results_curved}.
It can be observed that even for the \btssfullcons (and also the \btssfulldisp) variant, the analytical solution is not reproduced up to machine precision as the results show a non-vanishing stress state in the solid and a non-vanishing curvature in the beams.
However, these non-zero stress and curvature values, respectively, are introduced by the discretization error of the initial geometry, \ie the inability of the beam finite elements to exactly represent the curvature of the initial geometry, and are orders of magnitude smaller than the discretization errors associated with deformation states in typical application scenarios (and the error introduced by the \btssfullref variant).
It can be concluded that the discretization error for arbitrarily curved beam centerlines within the \btssfullcons and \btssfulldisp methods can be neglected as compared to the overall discretization error.

\subsection{Half-pipe with helix-shaped beam}
\label{sec:examples:coupling}

\begin{figure}
\centering
\includegraphics[page=1,scale=1]{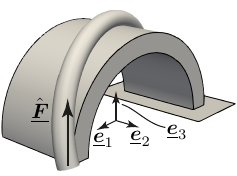}
\hfill
\includegraphics[page=1,scale=1]{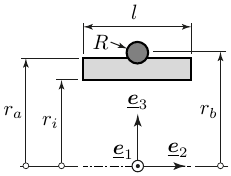}
\caption{%
Half-pipe with helix-shaped beam -- problem setup.
The left figure shows a 3D view of the problem and the right figure shows a cut through the $\ey-\ez$.
}
\label{fig:examples:coupling}
\end{figure}

In this example, a helix-shaped beam is coupled to the outer surface of a solid half-pipe, \cf \Cref{fig:examples:coupling}.
This example is introduced to further compare the three surface coupling types discussed in \Cref{sec:variants,sec:discretization:mortar_coupling}.
The solid half-pipe with length $l =  \unit[1]{m}$ has an outer radius $r_a = \unit[1]{m}$ and an inner radius $r_i = \unit[0.8]{m}$.
The pipe is modeled using a compressible Neo-Hookean material law ($\Esolid = \unit[1]{N/m^2}$, $\nusolid = 0$).
The solid is coupled to a helix-shaped beam with a radius~$r_b = \unit[1.05]{m}$ and a pitch of $\unit[2]{m}$.
The beam has a \cs radius $\radius = \unit[0.1]{m}$, Young's modulus $\Ebeam=\unit[50]{N/m^2}$ and Possion's ratio $\nubeam = 0$.
With the chosen geometric dimensions, the beam centerline does not exactly lie on the outer surface of the solid half-pipe, but is offset by a normal distance of $\unit[0.05]{m}$.
On one side of the half-pipe, a concentrated force $\hat{\tns{F}} = \unit[0.0004]{N} \ez$ is applied to the tip of the beam.
On the other side, the solid is fixed in all spatial directions.

\begin{figure}
\centering
\includegraphicsdpi{300}{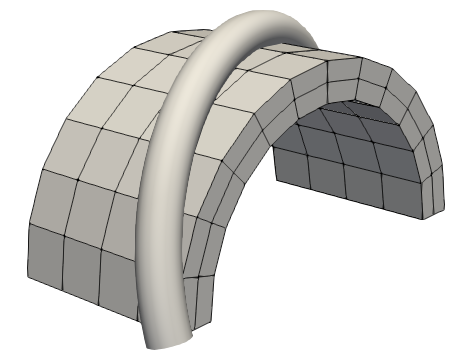}
\hfil
\includegraphicsdpi{300}{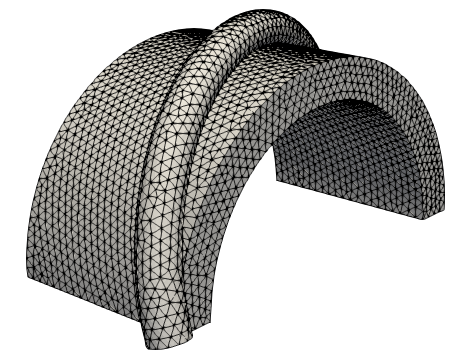}
\caption{Half-pipe with helix-shaped beam -- finite element discretization of the \btsso mesh (left) and full 3D mesh (right).}
\label{fig:examples:coupling_mesh}
\end{figure}

Coupling between the beam and the solid is realized with the three \btssfull coupling variants ($\penPos = \unit[10]{N/m^2}$, $\penRot = \unit[1]{Nm/m}$).
First-order Lagrange polynomials are employed to discretize both the positional and the rotational Lagrange multipliers.
The pipe is modeled with $2 \times 12 \times 4$ finite elements in radial, tangential and $\ey$ direction, respectively.
Eight-noded solid shell elements are employed, \cf\cite{Vu-Quoc2003a}.
It is important to note that solid shell elements are not shell elements in the classical sense.
They are eight-noded 3D hexahedral elements with enhanced assumed strain (EAS) and assumed natural strain (ANS) formulations to improve accuracy and numerical stability for thin structures.
Thus, the coupling scenario in this example can be classified as \btsso.
The beam is discretized using $10$ \sr beam finite elements.
The left part of \Cref{fig:examples:coupling_mesh} illustrates the finite element discretization of the \btsso model.
In the present example, the beam \cs{s} penetrate the solid coupling surface.
Therefore, it is also possible to discretize this example with a full 3D finite element mesh, where the beam itself is also modeled using 3D finite elements, \cf the right part of \Cref{fig:examples:coupling_mesh}.
The full 3D model is discretized with $50{,}480$ second-order tetrahedra (\tet{10}) elements.
Consequently, the full 3D model consists of $226{,}383$ degrees of freedom.
The discretization of the full 3D model has been chosen such that mesh convergence is guaranteed and it can be used as a reference solution to assess the quality of the results obtained with the three \btssfull variants.

\begin{figure*}
\centering
\subfigure[full 3D]{\label{fig:examples:coupling_initial_result:full}\includegraphicsdpi{300}{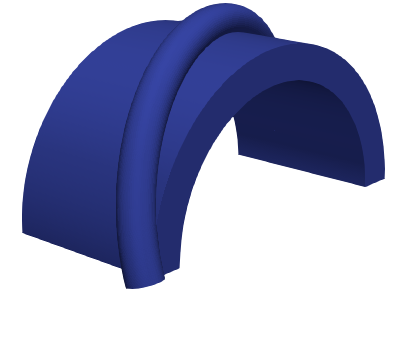}}
\hfil
\subfigure[\btssfullcons]{\label{fig:examples:coupling_initial_result:cons}\includegraphicsdpi{300}{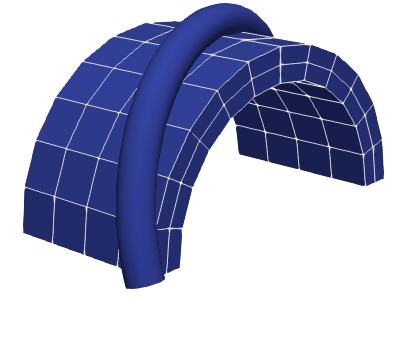}}
\hfil
\subfigure[\btssfullref]{\label{fig:examples:coupling_initial_result:ref}\includegraphicsdpi{300}{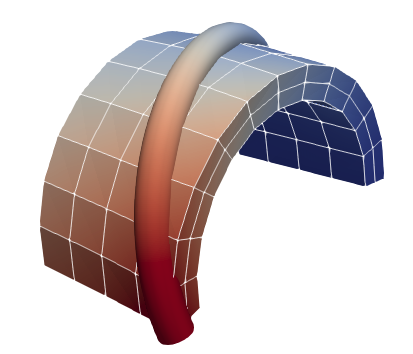}}
\hfil
\subfigure[\btssfulldisp]{\label{fig:examples:coupling_initial_result:disp}\includegraphicsdpi{300}{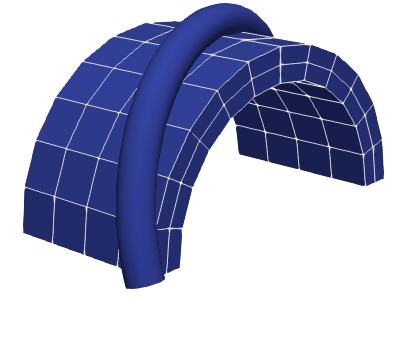}}
\hfil
{\normalsize\input{figures/examples_coupling_result_initial_color_bar.tex}}
\caption{Half-pipe with helix-shaped beam -- deformed configurations for the unloaded problem ($\hat{\tns{F}} = \tnsO$). The results for the various coupling schemes are shown and the contour plots visualize the displacement magnitude.}
\label{fig:examples:coupling_initial_result}
\end{figure*}

\begin{figure*}
\centering
\subfigure[full 3D]{\label{fig:examples:coupling_final_result:full}\includegraphicsdpi{300}{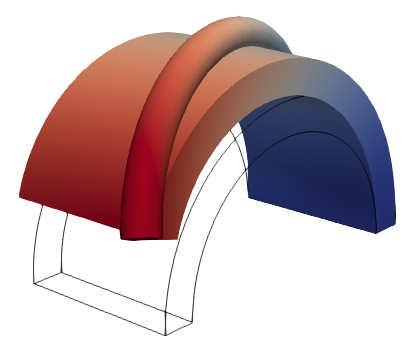}}
\hfil
\subfigure[\btssfullcons]{\label{fig:examples:coupling_final_result:cons}\includegraphicsdpi{300}{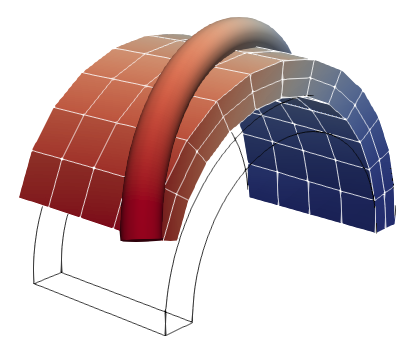}}
\hfil
\subfigure[\btssfullref]{\label{fig:examples:coupling_final_result:ref}\includegraphicsdpi{300}{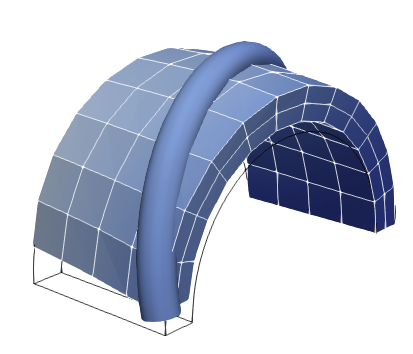}}
\hfil
\subfigure[\btssfulldisp]{\label{fig:examples:coupling_final_result:disp}\includegraphicsdpi{300}{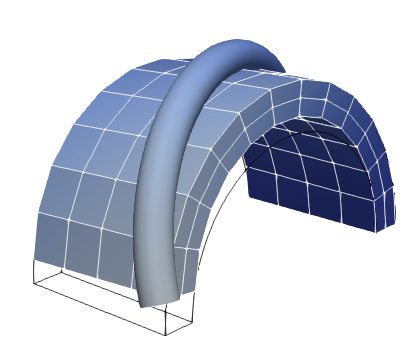}}
\hfil
{\normalsize\input{figures/examples_coupling_result_final_color_bar.tex}}
\caption{Half-pipe with helix-shaped beam -- deformed configurations for the loaded problem. The results for the various coupling schemes are shown and the contour plots visualize the displacement magnitude.}
\label{fig:examples:coupling_final_result}
\end{figure*}

\Cref{fig:examples:coupling_initial_result} visualizes the deformed configurations for the unloaded ($\hat{\tns{F}}=\tnsO$) structure.
Since no pre-stressing or prescribed initial deformations are applied to the structure, the analytical displacement field for the unloaded structure vanishes.
The \btssfullref variant exhibits non-vanishing displacements, \cf \Cref{fig:examples:coupling_initial_result:ref}.
This is because the coupling constraints in the reference configuration are only fulfilled by the \btssfullref variant \eqref{eq:problem_formulation:position_coupling_ref} if the beam centerline lies exactly on the surface, which is not the case in this example.
The coupling conditions thus force the beam centerline to lie on the surface.
This in turn leads to an artificial pre-stressing of the system as both the beam and the solid are deformed in order to fulfill the coupling constraints in the load-free reference configuration.
All other \btssfull coupling variants and the full 3D solution exhibit vanishing displacements up to machine precision as expected.
A quantitative comparison of the variants is given in \Cref{tbl:examples:coupling_initial_resuls}.
As discussed above, only the \btssfullref variant has a non-zero internal elastic energy $\Pi_{\nameinternal}$ (including the penalty coupling potential) and beam tip displacement~$\ubeam$ for the load-free state.
\Cref{fig:examples:coupling_final_result} visualizes the deformed configurations for the loaded structure.
It can be seen that the \btssfullcons variant closely resembles the full 3D reference solution.
The two other variants, \btssfullref and \btssfulldisp, exhibit a different solution than the full 3D model.
Again, quantitative comparisons of the variants are given in \Cref{tbl:examples:coupling_final_resuls}.
The results for the \btssfullref and \btssfulldisp show a large discrepancy with respect to the reference solution.
For the \btssfullref variant, this can easily be explained since already the initial (load-free) configuration does not match the reference solution.
For the \btssfulldisp variant, this discrepancy illustrates that the simplified coupling conditions are not able to accurately describe the coupling between the beam and the surface if the discretized beam centerline does not exactly lie within the discretized surface in the reference configuration.
Furthermore, the balance of internal and external moments around the origin shows that the conservation of angular momentum is not fulfilled by the \btssfulldisp variant.
Finally, the internal elastic energy and the beam tip displacement obtained with the \btssfullcons variant are very close to the reference solution, which is a remarkable feature considering the much simpler spatial discretization of the mixed-dimensional problem.
\begin{table*}
\centering
\caption[Half-pipe with helix-shaped beam -- numerical results for the unloaded problem]{Half-pipe with helix-shaped beam -- numerical results for the unloaded problem ($\hat{\tns{F}} = \tnsO$). The total internal elastic energy (including penalty coupling energy) $\Pi_\nameinternal$ and the beam tip displacement $\ubeam$ are stated.}
\label{tbl:examples:coupling_initial_resuls}
\begin{tabular}{lrrrrr}
\toprule
coupling type     & $\Pi_\nameinternal$ in $\unit{J} \cdot 10^{-4}$ &
$\ubeam$ in $\unit{m}$ \\
\midrule
full 3D & 0.00000 & [ 0.0,  0.0,  0.0 ] \\
\btssfullcons & 0.00000 & [ 0.0,  0.0, 0.0 ] \\
\btssfullref & 3.37499 & [ 0.24411,  -0.37493,  -0.03631 ] \\
\btssfulldisp & 0.00000 & [ 0.0,  0.0,  0.0 ] \\
\bottomrule
\end{tabular}
\end{table*}
\begin{table*}
\centering
\caption{Half-pipe with helix-shaped beam -- numerical results for the total internal elastic energy (including penalty coupling energy) $\Pi_\nameinternal$ and the beam tip displacement $\tns{u}$ are stated as well as their relative error.
Furthermore, the scaled sum of internal and external moments is stated.}
\label{tbl:examples:coupling_final_resuls}
\begin{tabular}{lrrrrr}
\toprule
coupling type     & $\Pi_\nameinternal$ in $\unit{J} \cdot 10^{-4}$ &   $\relError{\Pi_{\nameinternal}}{\Pi_{\nameinternal,\text{full3D}}}$ &
$\tns{u}$ in $\unit{m}$ & $\relError{\tns{u}}{\tns{u}_{\text{full3D}}}$ & $\frac{\norm{\Sigma\br{\tns{M}_\nameinternal + \tns{M}_\nameexternal}}}{\norm{\hat{\tns{F}}}l}$ \\
\midrule
full 3D & 1.14109 & -- & [ -0.08411,  0.55495,  -0.00476 ] & -- & 0.0000 \\
\btssfullcons & 1.12581 & 1.3392\% & [ -0.08077,  0.54627,  -0.00883 ] & 1.8088\% & 0.0000 \\
\btssfullref & 4.39811 & 285.4301\% & [ 0.05225,  0.10224,  -0.02442 ] & 84.3041\% & 0.0000 \\
\btssfulldisp & 4.74429 & 58.4232\% & [ -0.03799,  0.22497,  0.05944 ] & 60.4513\% & 126.5426 \\
\bottomrule
\end{tabular}
\end{table*}

Summing up, after the first two examples we can state that both presented simplifications of the \btssfull conditions, \btssfullref and \btssfulldisp, are not suitable for general purpose \btss coupling problems.
Only the \btssfullcons variant with a consistent handling of the surface normal vector, and its derivatives, passes the constant stress transfer tests and gives accurate results for more general loading conditions.
Therefore, only this consistent variant will be used in the remainder of this contribution to model the positional coupling between beam and surface.

\subsection{Supported structure}
\label{sec:examples:shell}

\begin{figure}
\centering
\includegraphics[scale=1]{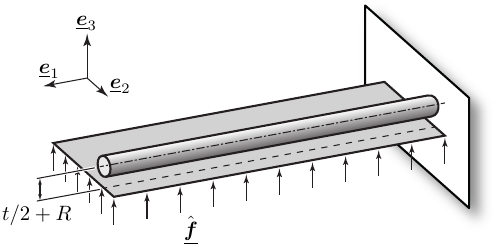}
\caption{%
Supported shell -- problem setup.
}
\label{fig:examples:btssc_rotation_plate:problem}
\end{figure}

In this example the importance of coupling both positions \emph{and} rotations within the \btss coupling scheme is demonstrated.
This is achieved by comparing the \btssfullcons (including rotational coupling) and \nameConsistent (without rotational coupling) schemes to each other.
Furthermore, we show the straightforward applicability of the proposed coupling method to connect 1D beams to 2D structural isogeometric shell formulations, \ie \btssh scenario.
The problem consists of a shell-like structure and a straight beam serving as a strut, \cf \Cref{fig:examples:btssc_rotation_plate:problem}.
The shell measures $\unit[3]{m} \times \unit[1]{m}$ with a thickness $\thickness = \unit[0.05]{m}$ and is loaded with a body load~$\hat{\tns{f}} = \unit[0.0002]{N/m^2} \ez$.
We employ a \kl shell formulation to model the shell, \cf\cite{Bischoff2004,Kiendl2009}, with material parameters set to $\Esolid = \unit[10]{N/m^2}$ and $\nusolid = 0$.
The shell is reinforced by a straight beam with circular \cs ($\radius=\unit[0.075]{m}$, $\Ebeam=\unit[100]{N/m^2}$ and $\nubeam = 0$).
The beam centerline is parallel to the $\ex$ axis and offset from the shell mid-surface by a distance of in surface normal direction, \ie the beam \cs exactly touches the top surface of the shell.
In~$\ey$ direction, the beam centerline is offset by a distance of~$\unit[0.35]{m}$ with respect to the middle of the shell.
At the right end, both and beam are fully clamped.
Apart from that, no displacement boundary conditions are applied to the system.

\begin{figure}
\centering
\includegraphics[scale=1]{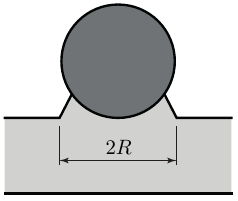}
\caption{Supported shell -- modeled weld line between the beam and the shell in the full 3D reference solution.}
\label{fig:examples:btssc_rotation_plate:weld}
\end{figure}

A full 3D reference solution is computed, where the shell as well as the beam are fully resolved with 3D solid finite elements. 
The material law for this reference solution is of compressible Neo-Hookean type and utilizes the same material parameters as employed within the shell formulation.
In this reference solution, the connection between the beam and the shell, \ie the weld line, has to be modeled.
\Cref{fig:examples:btssc_rotation_plate:weld} shows the fully resolved connection (weld line) between the beam and the shell which has a total width of~$2R$.
The weld line between beam and solid is assumed to be made up of the solid material.
The full model is discretized with first-order hexahedral (\hex{8}) elements, thus resulting in roughly $1{,}160{,}000$ elements and $1{,}250{,}000$ degrees of freedom to obtain mesh convergence.

In the 1D-2D \btss coupling problem, the coupling between the beam and the shell is realized with first-order Lagrange polynomials as shape functions for the positional and the rotational Lagrange multipliers ($\penPos = \unit[100]{N/m^2}$, $\penRot = \unit[0.5]{Nm/m}$).
The shell is modeled with~$30 \times 10$ $C^1$-continuous isogeometric shell elements (based on second-order \nurbs), \cf \cite{Kiendl2009}.
The beam is discretized using 10 \sr beam finite elements.
The total number of degrees of freedom for the \btss coupling problem is only.

\begin{figure*}
\centering
\subfigure[]{\label{fig:examples:btssc_rotation_plate:results_deformed:full_large}\includegraphicsdpi{300}{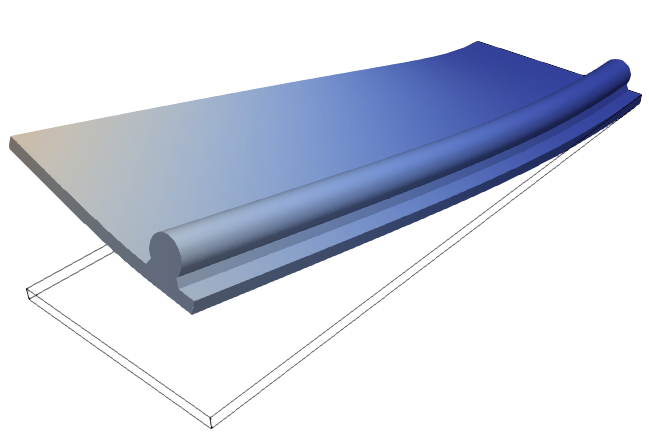}}
\hfil
\subfigure[]{\label{fig:examples:btssc_rotation_plate:results_deformed:bts_full_shell}\includegraphicsdpi{300}{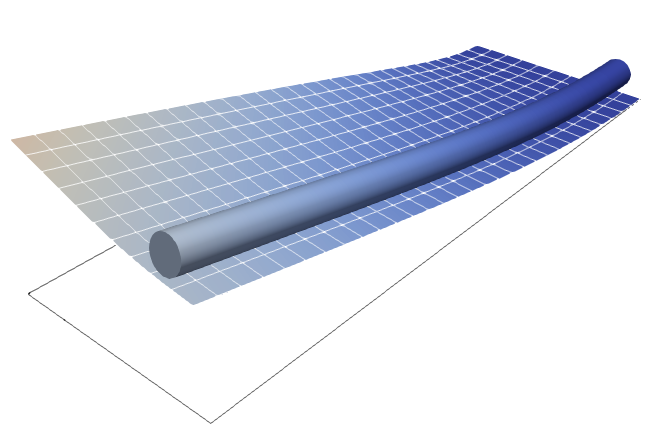}}
\hfil
\subfigure[]{\label{fig:examples:btssc_rotation_plate:results_deformed:bts_trans_shell}\includegraphicsdpi{300}{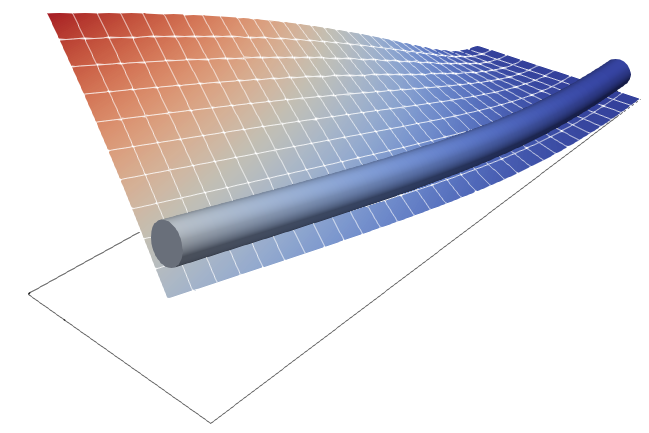}}
\\
{\normalsize\input{figures/examples_plate_rotation_color_bar.tex}}
\caption{Supported shell -- deformed configurations for various modeling techniques. \subref{fig:examples:btssc_rotation_plate:results_deformed:full_large} Full 3D model, \subref{fig:examples:btssc_rotation_plate:results_deformed:bts_full_shell} \btssfullcons (with rotational coupling) and \subref{fig:examples:btssc_rotation_plate:results_deformed:bts_trans_shell} \nameConsistent (without rotational coupling). The contour plots visualize the displacement magnitude.}
\label{fig:examples:btssc_rotation_plate:results_deformed}
\end{figure*}

\Cref{fig:examples:btssc_rotation_plate:results_deformed} visualizes the deformed configurations for the various models 
The full 3D reference solution as well as the 1D-2D \btssfullcons solution including rotational coupling behave very similarly, \ie the shell is bent upwards and the strut stiffens the shell, \cf \Cref{fig:examples:btssc_rotation_plate:results_deformed:full_large,fig:examples:btssc_rotation_plate:results_deformed:bts_full_shell}.
The \nameConsistent variant without rotational coupling, however, exhibits much larger deformations.
In that case, the rotational movement of the shell is not coupled to the rotations of the supporting beam, \ie the torsional stiffness of the beam is not directly coupled to the shell, thus resulting in an overall softer structural behavior, \cf \Cref{fig:examples:btssc_rotation_plate:results_deformed:bts_trans_shell}.
This clearly underlines the importance of including rotational coupling for \btss coupling problems to fully capture all relevant stiffening effects.
A more detailed comparison of the variants is given in \Cref{fig:examples:btssc_rotation_plate:results_plot}, where the configurations of the shell's left end are visualized.
Now it also becomes clear quantitatively that the displacement results obtained with the \nameConsistent variant without rotational coupling are unphysical due to the underestimated overall stiffness of the structure.
Furthermore, the full 3D model and the \btssfullcons model exhibit a very good agreement of the resulting displacement curves.
Considering that the latter variant reduces the number of degrees of freedom by a factor of about 1000, this is a remarkable result and showcases the efficiency of the \btssfullcons coupling method for reinforced shell applications.
\begin{figure*}
\centering
{\normalsize\begin{tikzpicture}
\begin{axis}[width=6.cm, height=3.7082039324993685cm, scale only axis, xlabel=$\underline{\boldsymbol{e}}_2$, ylabel=$\underline{\boldsymbol{e}}_3$, legend pos=outer north east, legend columns=1, legend cell align={left}, legend style={/tikz/every even column/.append style={column sep=0.3cm}}, yticklabel style={anchor=east,/pgf/number format/precision=2,/pgf/number format/fixed,}, unit vector ratio={1 1}]
\addplot+[, mark repeat=17] table{figures/data/examples_plate_rotation_plot_data_1.dat};
\addlegendentry{full 3D}
\addplot+[, mark repeat=20] table{figures/data/examples_plate_rotation_plot_data_2.dat};
\addlegendentry{\btssfullcons}
\addplot+[, mark repeat=20] table{figures/data/examples_plate_rotation_plot_data_3.dat};
\addlegendentry{\nameConsistent}
\end{axis}
\end{tikzpicture}}
\caption{Supported shell -- deformed configurations of the shell's left end for various modeling techniques.}
\label{fig:examples:btssc_rotation_plate:results_plot}
\end{figure*}

\subsection{Towards biomedical applications}
\label{sec:examples:artery}

The last example is designed to give an outlook towards real-life applications and the suitability of the proposed \btssfull approach for more complex coupling scenarios.
Specifically, we want to analyze the applicability of our approach in the context of vascular angioplasty.
To this end, we set up a variant of the well-known fluid-structure interaction (FSI) benchmark problem of a pressure wave traveling through an elastic tube, that was originally proposed in \cite{Gerbeau2003} to validate the suitability of FSI algorithms for blood flow simulations.
In addition to the original problem, we will use our \btssfull coupling approach to capture the effect of a diamond-shaped stent structure on the behavior of the overall system.
In particular, we expect to capture the large change in compliance between the stented and unstented regions of the pipe, thus leading to stress peaks in these transitional regions as well as an altered fluid flow. Such effects have been linked to the occurrence of in-stent restenosis and are of high interest when analyzing the suitability of endovascular devices and their effect on the patient \cite{Kim2013, Colombo2021, Kohler1992}.

\begin{figure*}
    \centering
    \subfigure[]{\label{fig:examples:fsi_geometry:artery}\includegraphics[scale=1]{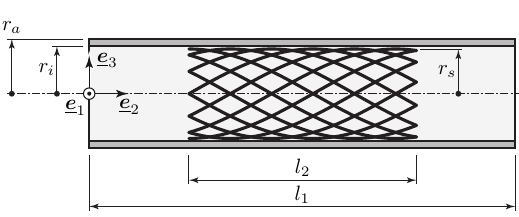}}
    \hfil
    \subfigure[]{\label{fig:examples:fsi_geometry:stent}\includegraphics[scale=1]{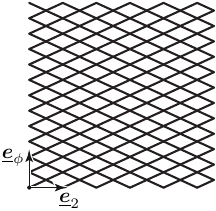}}
    \caption{Towards biomedical applications -- problem setup. \subref{fig:examples:fsi_geometry:artery} Geometric configuration of the stent and
    artery, and \subref{fig:examples:fsi_geometry:stent} unwrapped stent geometry.}
    \label{fig:examples:fsi_geometry}
\end{figure*}

\begin{table}
\begin{center}
\caption{Table containing the parameters for the stented elastic pipe problem.}
\begin{tabular}{llr}
\toprule
	Geometry & $r_i$           &         $\unit[0.0125]{m}$ \\
	         & $r_a$           &        $\unit[0.01375]{m}$ \\
	         & $r_s$           &        $\unit[0.01246]{m}$ \\
	         & $l_1$           &           $\unit[0.15]{m}$ \\
	         & $l_2$           &           $\unit[0.06]{m}$ \\
\midrule
	Beam     & $E_B$           & $\unit[9\cdot10^8]{N/m^2}$ \\
	         & $\rho_B$        &     $\unit[7800]{kg/m^3} $ \\
	         & $\nu_B$         &                      $0.3$ \\
	         & $R$             &         $\unit[0.0004]{m}$ \\
\midrule
	structure    & $E_S$           & $\unit[3\cdot10^5]{N/m^2}$ \\
	         & $\rho_S$        &     $\unit[1200]{kg/m^3} $ \\
	         & $\nu_S$         &                      $0.3$ \\
\midrule
	Fluid    & $p_{\text{in}}$ &        $\unit[500]{N/m^2}$ \\
	         & $\rho_F$        &     $\unit[1000]{kg/m^3} $ \\
	         & $ \eta_F$       &   $\unit[0.003]{kg/(m s)}$ \\
\bottomrule
\end{tabular}
\end{center}
\label{tab:fsi_parameters}
\end{table}

As in the original benchmark problem, a constant pulse~$p_\text{in}$ is applied for $\unit[3\cdot10^{-3}]{s}$ at the fluid inlet. Besides the pulse, zero traction conditions are applied to the fluid inflow as well as outflow boundary on the left and right end of the pipe, respectively, while both ends of the pipe are assumed to be clamped.
In addition to the \btssfull problem introduced in \Cref{sec:problem_formulation}, this example contains a fluid, modeled as Newtonian with a constant dynamic viscosity $\eta_F$ and a density~$\rho_F$, using the incompressible Navier-Stokes equations. \Cref{fig:examples:fsi_geometry:artery} illustrates the problem setup.
The fluid is coupled to the structure via classical surface-coupled FSI \cite{Kloeppel2011} in a partitioned manner aided by a matrix-free Newton Krylov method \cite{Kuettler2008} to accelerate convergence. Classical no-slip conditions are enforced on the FSI boundary.
The beam centerline geometry depicted in Figure \ref{fig:examples:fsi_geometry:stent} is wrapped around a cylinder with a radius of $r_s = r_i - R$ to create the used diamond-shaped stent geometry such that the stent perfectly fits into the pipe structure up to an offset the size of the beam radius.
Since FSI problems are necessarily transient, the \btssfull problem is enhanced by a Generalized-$\alpha$ Lie group time integration method for all structural degrees of freedom \cite{Bruels2010, Bruels2012}. Here, the parameters are chosen to obtain a fully implicit scheme, and a time step size $\Delta t=\unit[10^{-4}]{s}$ is used. To the fluid field, a classical second-order accurate Generalized-$\alpha$ time integration scheme, with the same time step size as for the structure field, is applied \cite{Jansen2000}.
The mortar-type \btssfullcons method is used with linear shape functions for the Lagrange multiplier fields and the penalty parameters $\penPos=\unit[10^9]{N/m^2}$ and $\penRot=\unit[10^{-1}]{Nm/m}$. To discretize the problem, $264$ \sr beam elements, $2{,}880$ solid shell elements and $22{,}800$ PSPG/SUPG stabilized Q1-Q1 fluid elements with an additional div-grad stabilization term~\cite{schott2015} are employed.
All dimensions and material parameters of the problem setup are summarized in Table \ref{tab:fsi_parameters}.

\begin{figure*}
\centering
\begin{tikzpicture}[baseline={([yshift=27pt]current bounding box.south)}]
\node (image0) at (0,0) {\includegraphics[scale=0.24]{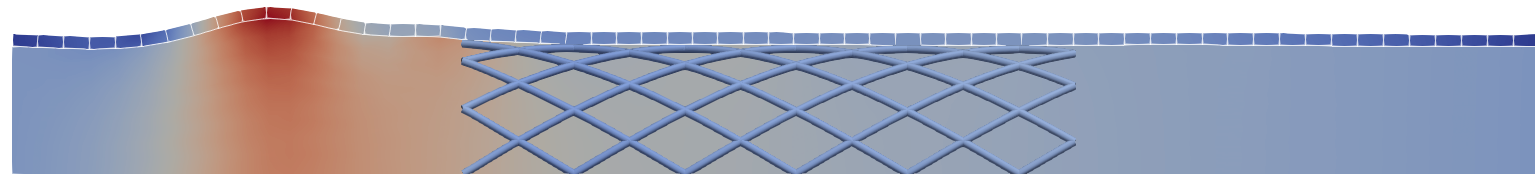}};
\node[anchor=north] (image1) at ([yshift=-2mm]image0.south) {\includegraphics[scale=0.24]{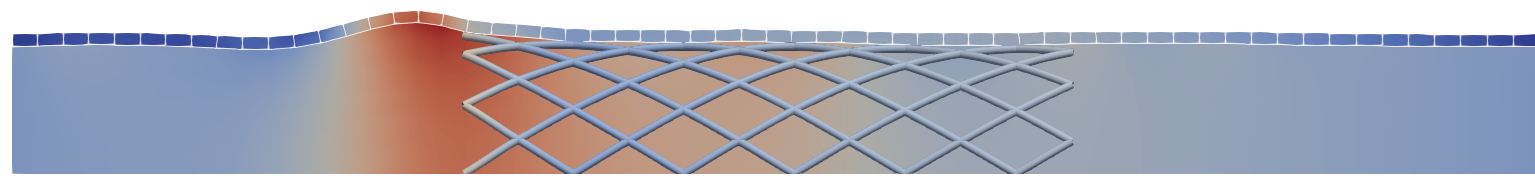}};
\node[anchor=north] (image2) at ([yshift=-2mm]image1.south) {\includegraphics[scale=0.24]{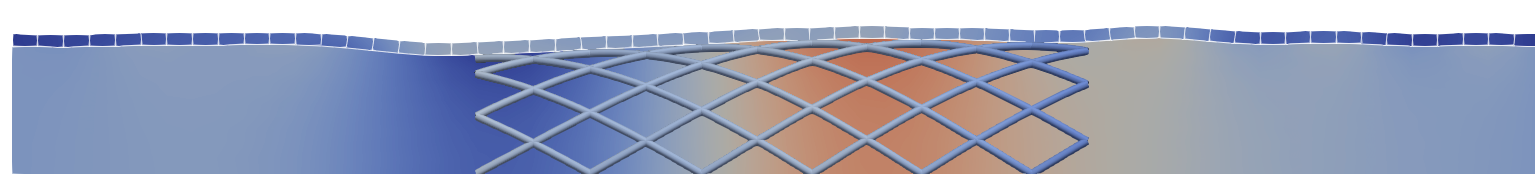}};
\node[anchor=north] (image3) at ([yshift=-2mm]image2.south) {\includegraphics[scale=0.24]{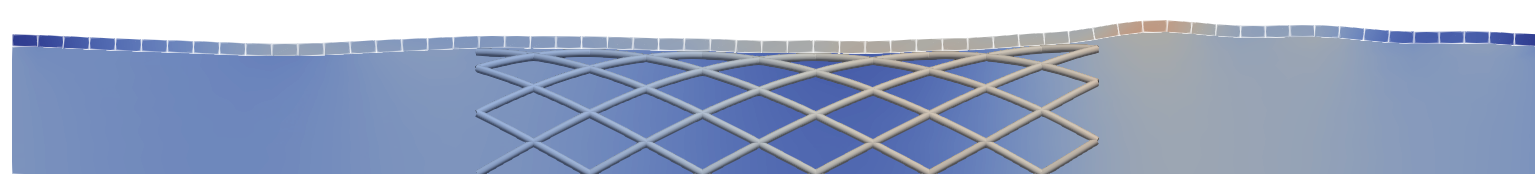}};
\node[anchor=east] at ([yshift=2mm]image0.west) {\footnotesize(a)};
\node[anchor=east] at ([yshift=2mm]image1.west) {\footnotesize(b)};
\node[anchor=east] at ([yshift=2mm]image2.west) {\footnotesize(c)};
\node[anchor=east] at ([yshift=2mm]image3.west) {\footnotesize(d)};
\end{tikzpicture}
{\normalsize\input{figures/examples_fsi_deformed_legend.tex}}
\caption{Towards biomedical applications -- deformed configuration of the stented elastic pipe problem at various simulation times. The snapshots are taken at (a) $t=\unit[0.01]{s}$, (b) $t=\unit[0.016]{s}$, (c) $t=\unit[0.024]{s}$ and (d) $t=\unit[0.030]{s}$ respectively.
The norm of the displacements is shown in the structure and the pressure is shown in the fluid.
The displacements are scaled with a factor of 15.}
\label{fig:examples:fsi:disp}
\end{figure*}

Figures \ref{fig:examples:fsi:disp}(a) to \ref{fig:examples:fsi:disp}(d) depict the structural displacement scaled with a factor of 15 and the fluid pressure after $\unit[0.01]{s}$, $\unit[0.016]{s}$, $\unit[0.024]{s}$ and $\unit[0.030]{s}$.
It is evident, that the wall displacement caused by the pressure wave in the stiffer stented region in \Cref{fig:examples:fsi:disp} is smaller than in the unstented region.
This, in turn, affects the fluid since a constant flow throughput requires increased velocities within the stented region compared to the more compliant unstented region.
\Cref{fig:examples:fsi:plots} illustrates the fluid velocity $v_2$ in channel direction along the pipe's centerline.
The fluid velocity plot demonstrates the previously mentioned phenomenon as the maximum fluid velocity increases slightly and the wave broadens while traveling through the stented region.
This effect on the fluid flow is still visible even after the pressure wave leaves the stented region.
\begin{figure*}
\centering
{\normalsize\begin{tikzpicture}
\begin{axis}[width=6.cm, height=3.7082039324993685cm, scale only axis, xlabel=$\ey$, ylabel=$v_2$, legend pos=outer north east, legend cell align={left}]
\addplot+[mark=none] table{figures/data/examples_fsi_stent_velocity_data_1.dat};
\addplot+[mark=none] table{figures/data/examples_fsi_stent_velocity_data_2.dat};
\addplot+[mark=none] table{figures/data/examples_fsi_stent_velocity_data_3.dat};
\addplot+[mark=none] table{figures/data/examples_fsi_stent_velocity_data_4.dat};
\addlegendentry{\unit[0.010]{s}}
\addlegendentry{\unit[0.016]{s}}
\addlegendentry{\unit[0.024]{s}}
\addlegendentry{\unit[0.030]{s}}
\draw[shorten >=0pt,shorten <=2pt, gray] (-0.10,0) -- (0.25,0);
\end{axis}
\end{tikzpicture}}
\caption{Plots of the fluid velocity $v_2$ in channel direction along the pipe's centerline.}
\label{fig:examples:fsi:plots}
\end{figure*}

\begin{figure*}
\centering
\begin{minipage}{0.8\textwidth}
\subfigure[]{\label{fig:examples:fsi:lambda:1}\includegraphicsdpi{300}{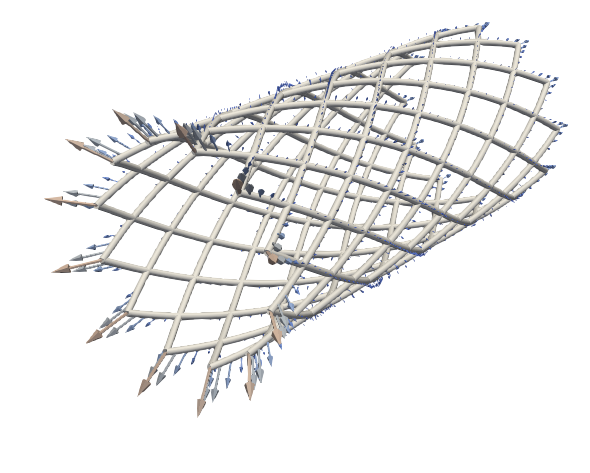}}
\hfil
\subfigure[]{\label{fig:examples:fsi:lambda:2}\includegraphicsdpi{300}{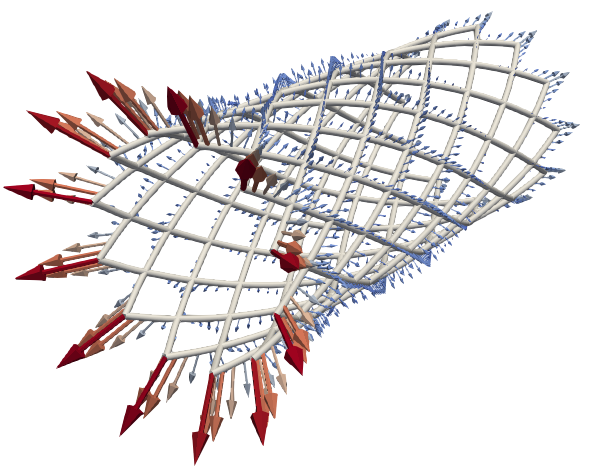}}
\\
\subfigure[]{\label{fig:examples:fsi:lambda:3}\includegraphicsdpi{300}{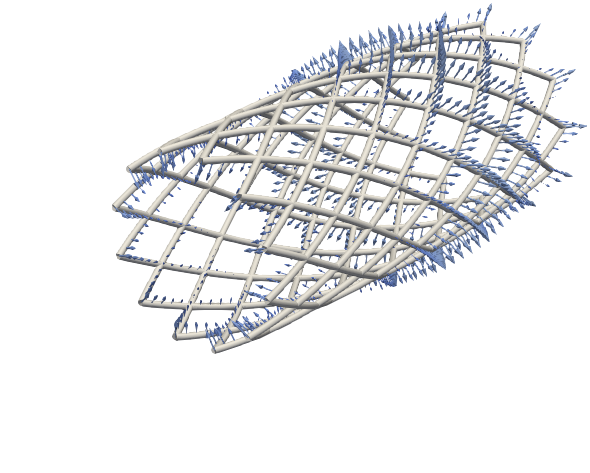}}
\hfil
\subfigure[]{\label{fig:examples:fsi:lambda:4}\includegraphicsdpi{300}{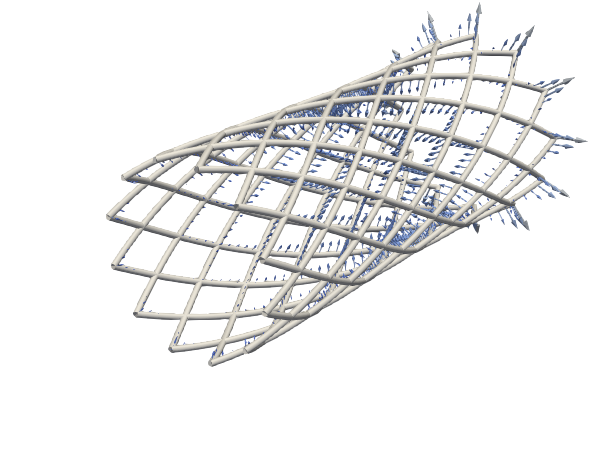}}
\end{minipage}
{\normalsize\input{figures/examples_fsi_lambda_legend.tex}}
\caption{Negative coupling line loads for \btssLong coupling at various simulation times -- the snapshots are taken at (a) $t=\unit[0.01]{s}$, (b) $t=\unit[0.016]{s}$, (c) $t=\unit[0.024]{s}$ and (d) $t=\unit[0.030]{s}$ respectively. Only the positional Lagrange multiplier field is shown and five values are visualized along each beam element. The displacements are scaled with a factor of 15.}
\label{fig:examples:fsi:lambda}
\end{figure*}

While the change of compliance in the artery, and thus also its effect on the fluid flow, could also be modeled by a simpler homogenized approach, the proposed approach allows to quantify the forces interchanged on the coupling interface.
\Cref{fig:examples:fsi:lambda:1,fig:examples:fsi:lambda:4} illustrate the coupling interactions, \ie the line loads excerted on the beam system by the surface.
In general, it can be observed that the interaction is highest at the ends of the stent, \ie at the transition between a compliant and a very stiff region.
This is particularly notable in Figure \ref{fig:examples:fsi:lambda:2}, where the pressure wave is right at the transition between the unstented and stented region.
Furthermore, dividing the 1D coupling loads by the beam diameter results in an approximation of the interaction stresses between the beam and the artery.
The maximum absolute values of normal and shear stresses can be estimated for this example as $\unit[1.8179 \cdot 10^3]{N/m^2}$ and $\unit[1.28899 \cdot 10^3]{N/m^2}$ (not visualized in the figures), respectively.

The demonstrated example certainly represents a simplified model. In particular, the use of \btssLong coupling, as presented here, instead of frictional \btss contact, prevents the observation of some real-life phenomena such as stent migration.
Nevertheless, because of growth and remodeling of the artery and successive protrusion of the stent struts, coupling, \ie mesh tying, is a valid assumption in many patient-specific cases.
A further interesting novel computational method to incorporate was recently reported in \cite{Hagmeyer2022, Hagmeyer2024}.
It enables capturing the effect of the stent struts on the fluid flow, which is linked to altered wall shear stresses that may lead to in-stent restenosis \cite{Johari2020, Pant2010}.
In any case, the presented simulation results serve as a proof of concept to show that the proposed \btssfull coupling approach can generally be used for geometrically complex beam systems such as stent geometries. The ability to capture important phenomena, such as changes in compliance and its effect on the blood flow as well as the distribution of the interaction forces, which may provide insight into the long-term success of vascular angioplasty, has been demonstrated.

\section{Conclusion}

In this work, we have proposed a 1D-2D mixed-dimensional coupling method to consistently couple 1D \cosserat beams to 2D surfaces.
We consider both 2D surfaces of classical 3D \boltz continua and reduced dimensional 2D shell formulations.
In the presented coupling methods, six coupling constraints act along the beam centerline, \ie three positional constraints and three rotational constraints.
Three different variants of the positional coupling constraints have been investigated.
One of them, the \emph{consistent} variant, requires the expensive evaluation of the current surface normal field.
The other two variants are commonly used in \sutsu mesh tying problems.
Numerical examples show that only the consistent positional coupling constraints, \ie with inclusion of the surface normal vector, lead to physically correct results and fulfill basic mechanical consistency properties, such as conservation of angular momentum.
These findings can also be transferred to classical \sutsu coupling problems, where the continuous surfaces are non-matching.
Existing coupling methods for the rotational degrees of freedom are extended by constructing a suitable surface triad field.
The Lagrange multiplier method is used to enforce the positional and rotational coupling constraints.
The coupling equations are discretized using a mortar-type approach and the resulting discrete constraint equations are regularized via a weighted penalty approach.
Furthermore, the numerical examples illustrate the importance of combining both positional \emph{and} rotational coupling via a practically motivated example.
Finally, a multi-physics simulation, inspired by models of stented arteries, has demonstrated the method's suitability for complex beam geometries and its ability to capture global effects on the solid as well as the fluid field.

Future work will focus on the extension of the presented beam-to-surface coupling approach to beam-to-surface contact and finite sliding problems, \ie replacing the coupling constants with unilateral and frictional contact constraints.

\begin{acknowledgements}
Sketches in this work have been created using the Adobe Illustrator plug-in LaTeX2AI (\url{https://github.com/isteinbrecher/LaTeX2AI}).
\end{acknowledgements}

\appendix
\section{Limitations of beam-to-volume coupling applied to the \btsso scenario}

In the case of beams coupled to the 2D surface of 3D solids (\btsso scenario), one might consider evaluating a beam-to-volume coupling formulation, \eg \cite{Steinbrecher2020, Steinbrecher2022}, on the boundary of the volume (or even beyond).
While this approach may work in special cases, it generally leads to undesirable behavior.
This appendix examines these limitations and illustrates the potential pitfalls of applying beam-to-volume coupling to \btsso scenarios.

\subsection{Solid volume triad}
\label{sec:appendix_solid_surface_triad}

\begin{figure}
	\centering
	\includegraphics[page=1,scale=1]{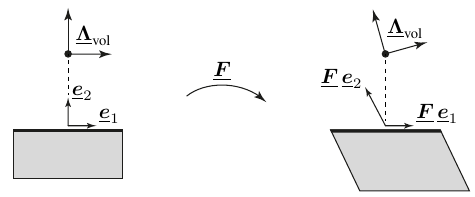}
	\caption{%
		Illustration of the influence of out-of-plane solid deformations on the solid deformation gradient at the surface.
		If the volume triad construction according to~\cite{Steinbrecher2022} is applied, the purely out-of-plane deformations influence the triad evaluated at the surface.
	}
	\label{fig:surface_triad:surface_kinematics}
\end{figure}

For the construction of the solid volume triads $\triadVolume$ in \cite{Steinbrecher2022} the solid deformation gradient $\F$ was used.
It has been shown that the rotation tensor obtained via a polar decomposition of the solid deformation gradient fulfills both required properties (i) and (ii) from~\Cref{sec:surface_triad}, and represents the solid material directors in an optimal manner.
Furthermore, a slightly modified construction of the volume triad was presented, which fixes an averaged solid material director to the volume triad and eliminates the need for the computationally expensive evaluation of the polar decomposition (and its second derivatives) at Gauss-point level.
In theory, when 2D surfaces of 3D volumes are considered, the solid volume triad definitions from~\cite{Steinbrecher2022} can also be employed for a \btss problem, where the solid deformation gradient is evaluated at the surface.
However, in this case the surface triad field does not only depend on the surface deformation, but also on the deformation inside the volume.
This is illustrated in~\Cref{fig:surface_triad:surface_kinematics}, where the solid exhibits deformations inside the solid volume, while the surface geometry stays the same.
The solid deformation gradient at the surface changes due to the deformation of material fibers inside the solid volume.
Thus, the purely out-of-plane deformations influence the triad evaluated at the surface.
However, from an intuitive physical point of view the \emph{orientation} of the surface does not change.
This is not the case with the surface triad construction presented in~\Cref{sec:surface_triad}.

\subsection{\BtsvcXFull}

\begin{figure}
	\centering
	\includegraphics[scale=1]{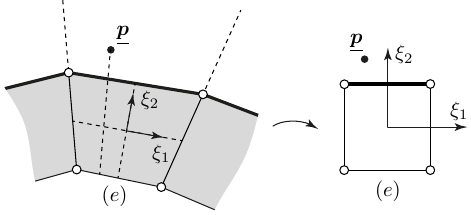}
	\caption{%
		Projection of point $\tns{p}$ to the parameter space of the solid finite element $\esolid$ with \btsvcX.
		For illustrative purposes a 2D example is shown.
	}
	\label{fig:discretization:xvolume}
\end{figure}

\begin{figure}
	\centering
	\subfigure[]{\label{fig:discretization:xvolume_bad_a}\includegraphics[page=1,scale=1]{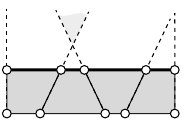}}
	\hfil
	\subfigure[]{\label{fig:discretization:xvolume_bad_b}\includegraphics[page=2,scale=1]{figures/discretization_xvolume_bad.pdf}}
	\caption{%
		Problematic cases for \btsvcX.
		\subref{fig:discretization:xvolume_bad_a} Distorted elements in negative normal direction of the solid surface.
		\subref{fig:discretization:xvolume_bad_b} General tetrahedral mesh.
		Gray areas indicate where a projection to the surface fails.
		For illustrative purposes a 2D example is shown.
	}
	\label{fig:discretization:xvolume_bad}
\end{figure}

The main difference between the proposed \nameConsistent coupling procedure, \cf \Cref{sec:variants_consisent}, and the \btsvc method, \cf \cite{Steinbrecher2020}, is a term accounting for the normal distance between the beam and the surface.
The discretization of this term introduces rather complex coupling terms, which require the evaluation of a surface normal field.
In the case of 2D surfaces of 3D solid volumes, an alternative to the \nameConsistent method is to use an extended version of the \btsvc scheme proposed in~\cite{Steinbrecher2020}, which shall be denoted as the \emph{extended} positional beam-to-volume coupling (\btsvcX) scheme in the following.
The idea of this \btsvcX scheme is to simply project points on the beam centerline to an extended solid parameter space, \ie projections that lie outside of the volume are still admissible.
Thus, no closest point projection with the surface normal field is required.
This is exemplarily illustrated in \Cref{fig:discretization:xvolume}.
The point $\tns{p}$ is projected to the parameter space of the solid finite element $(\esolidName)$, and, although the $\xi_2$ coordinate of the projection point lies outside of the solid finite element domain, the projection will still be used in the evaluation of~$\M$.
In this case, there are no coupling terms dependent on the surface normal distance.
At first glance this approach might seem very appealing as there is no need for evaluating the surface normal vector and its derivatives.
Furthermore, the same implementation as in \btsvc problems can be used.
However, there are two significant drawbacks of this approach: (i)~The projection of beam centerline points onto the surface is highly dependent on the solid finite element mesh.
\Cref{fig:discretization:xvolume_bad} illustrates cases where the \btsvcX method fails.
In \Cref{fig:discretization:xvolume_bad_a} the solid finite elements are distorted in negative normal direction of the coupling surface, such that a unique projection is not possible in the shaded areas.
(ii)~The \btsvcX method only works well for grid-like hexahedral meshes of the solid.
Unstructured hexahedral meshes or tetrahedral meshes lead to problems due to non-unique projections, \cf \Cref{fig:discretization:xvolume_bad_b}.
These drawbacks emphasize the importance of the presented \nameConsistent coupling scheme.

\bibliographystyle{spmpsci}      
\bibliography{literature_filtered,steinbrecher_filtered}

\end{document}